\documentclass[preprintnumbers,article,amsmath,amssymb,floatfix,10pt,prd,onecolumn,superscriptaddress,nofootinbib]{revtex4-2}

\usepackage{titlesec}
\usepackage[toc]{appendix}
\usepackage{caption}
\titleformat*{\section}{\LARGE\bfseries}
\titleformat*{\subsection}{\Large\bfseries}
\titleformat*{\subsubsection}{\large\bfseries}
\titleformat*{\paragraph}{\large\bfseries}
\titleformat*{\subparagraph}{\large\bfseries}
\usepackage{bm}
\usepackage{amsfonts}
\usepackage{latexsym}
\usepackage[utf8]{inputenc}
\usepackage{graphicx}
\usepackage{amsmath}
\usepackage{palatino}
\usepackage{mathpazo}
\usepackage{textcomp}
\usepackage{float}
\usepackage{booktabs}
\usepackage{dcolumn}
\usepackage{ragged2e}
\usepackage{hyperref}
\hypersetup{colorlinks,citecolor=blue}
\hypersetup{colorlinks=true,linkcolor=blue,filecolor=magenta,urlcolor=blue}
\usepackage{amsthm}
\usepackage{xcolor}
\usepackage{orcidlink}
\usepackage{epsfig}
\usepackage{commath}
\usepackage{cancel, ulem}
\usepackage{subcaption}
\usepackage{caption}
\usepackage{multirow}
\newcommand{\m}{\mathring}
\newcommand{\be}{\begin{equation}}
\newcommand{\ee}{\end{equation}}
\newcommand{\bea}{\begin{eqnarray}}
\newcommand{\eea}{\end{eqnarray}}
\newcommand{\eeas}{\end{eqnarray*}}
\newcommand{\beas}{\begin{eqnarray*}}
\usepackage{adjustbox}

\def\jnl@style{\it}
\def\aaref@jnl#1{{\jnl@style#1}}

\def\aaref@jnl#1{{\jnl@style#1}}

\def\aj{\aaref@jnl{AJ}}                   
\def\apj{\aaref@jnl{ApJ}}                 
\def\apjl{\aaref@jnl{ApJ}}                
\def\apjs{\aaref@jnl{ApJS}}               
\def\apss{\aaref@jnl{Ap\&SS}}             
\def\aap{\aaref@jnl{A\&A}}                
\def\aapr{\aaref@jnl{A\&A~Rev.}}          
\def\aaps{\aaref@jnl{A\&AS}}              
\def\mnras{\aaref@jnl{Mon.~Not.~Roy.~Astron.~Soc.}}             
\def\prd{\aaref@jnl{Phys.~Rev.~D}}        
\def\prc{\aaref@jnl{Phys.~Rev.~C}}  
\def\prl{\aaref@jnl{Phys.~Rev.~Lett.}}    
\def\qjras{\aaref@jnl{QJRAS}}             
\def\skytel{\aaref@jnl{S\&T}}             
\def\ssr{\aaref@jnl{Space~Sci.~Rev.}}     
\def\zap{\aaref@jnl{ZAp}}                 
\def\nat{\aaref@jnl{Nature}}              
\def\aplett{\aaref@jnl{Astrophys.~Lett.}} 
\def\apspr{\aaref@jnl{Astrophys.~Space~Phys.~Res.}} 
\def\physrep{\aaref@jnl{Phys.~Rep.}}      
\def\physscr{\aaref@jnl{Phys.~Scr}}       
\def\commat{\aaref@jnl{Comm.~Math.~Phys.}}              
\def\science{\aaref@jnl{Science}}               
\def\cqg{\aaref@jnl{Classical Quant.~Grav.}}            
\def\jpcs{\aaref@jnl{JPCS}}                                     
\def\ijmpd{\aaref@jnl{Int.~J.~Mod.~Phys.~D}}                    
\def\grg{\aaref@jnl{Gen.~Relat.~Gravit.}}               
\def\rpp{\aaref@jnl{Rep.~Prog.~Phys.}}          
\def\npa{\aaref@jnl{Nucl.~Phys.~A}}        
\def\lrr{\aaref@jnl{Living Rev.~Rel.}}                   
\def\jcap{\aaref@jnl{J.~Cosmology Astropart.~Phys.}}    
\def\rmp{\aaref@jnl{Rev.~Mod.~Phys.}}   
\def\epjc{\aaref@jnl{Eur.~Phys.~J.~C}} 
\def\plb{\aaref@jnl{~Phy.~Lett.~B}} 
\def\mpla{\aaref@jnl{Mod.~Phy.~Lett.~A}} 
\def\arxiv{\aaref@jnl{arxiv.org}}

\allowdisplaybreaks[1]
\renewcommand{\arraystretch}{1.1}
\begin{document}
\title{Scalar-Tensor Symmetric Teleparallel Gravity: Reconstruct the Cosmological History with a Steep Potential}

\author{Ghulam Murtaza\orcidlink{0009-0002-6086-7346}}
\email{ghulammurtaza@1utar.my}
\affiliation{Department of Mathematical and Actuarial Sciences, Universiti Tunku Abdul Rahman, Jalan Sungai Long,
43000 Cheras, Malaysia}
\author{Avik De\orcidlink{0000-0001-6475-3085}}
\email{avikde@um.edu.my}
\affiliation{Institute of Mathematical Sciences, Faculty of Science, Universiti Malaya, 50603 Kuala Lumpur, Malaysia}

\author{Andronikos Paliathanasis\orcidlink{0000-0002-9966-5517}}
\email{anpaliat@phys.uoa.gr}
\affiliation{Departamento de Matem\`{a}ticas, Universidad Cat\`{o}lica del 
Norte, Avda. Angamos 0610, Casilla 1280 Antofagasta, Chile}
\affiliation{Institute of Systems Science, Durban University of Technology, Durban 4000,
South Africa}
\affiliation{Centre for Space Research, North-West University, Potchefstroom 2520, South Africa}
\affiliation{National Institute for Theoretical and Computational Sciences (NITheCS), South Africa}

\footnotetext{GM acknowledges Universiti Tunku Abdul Rahman Research Fund project IPSR/RMC/UTARRF/2023
C1/A09. AD acknowledges Universiti Malaya BKP-ECRG Grant (Project No. BKP119-2025
ECRG). AP was partially supported by FONDECYT Grant 1240514}

\begin{abstract}
Within the framework of scalar-non-metricity gravity, we introduce a steep potential together with a power-law coupling function and investigate whether the acceleration phases of the universe can be consistently described by this model. In the symmetric teleparallel formulation, and under a Friedmann--Lemaître--Robertson--Walker background, three distinct branches of the connection arise, leading to three different cosmological scenarios. We perform a detailed dynamical analysis of these models by examining the phase space and determining the asymptotic cosmological solutions. The analysis reveals a rich hierarchy of critical points, including matter-dominated epochs, kinetic-dominated stiff-fluid regimes, and steep potential-dominated de Sitter solutions, along with asymptotic trajectories that approach Big Crunch or Big Rip singularities, as well as transient, unstable matter-dominated eras. The stability of the steep potential-dominated de Sitter points is further studied using Center Manifold Theory, showing that, under specific parametric conditions, the model can provide a unified description of both the early and late-time acceleration phases of the universe.

\end{abstract}
\maketitle
\tableofcontents
 
\section{Introduction}\label{sec0}
Observational evidence from the latest cosmological studies \cite{Riess, Tegmark, Kowalski, Komatsu, Abdalla, Lynch2024} indicates that the universe is undergoing an accelerated phase. To account for this acceleration within the framework of General Relativity (GR), an additional matter component with negative pressure, known as dark energy, must be introduced into the gravitational field equations \cite{Gorini:2004by,Ratra:1987rm,Brax:2017idh,Benaoum:2019kyj}. Nevertheless, cosmologists have been motivated to develop theories beyond GR in an effort to address the hidden aspects of the universe. In this approach, the dynamics responsible for cosmic acceleration are attributed to geometric degrees of freedom introduced in the modified Einstein–Hilbert Action, giving rise to modified or alternative theories of gravity \cite{Clifton2012}.

The simplest of these extensions is $f(\mathring{R})$ gravity, where $\mathring{R}$ denotes the Ricci scalar in the Einstein-Hilbert action \cite{Felice2010}. Scalar-tensor theories \cite{Brans:1961sx,Faraoni:2004pi,Quiros:2019ktw} form another important class of extensions and are regarded as prime candidates for explaining late-time cosmic acceleration without invoking exotic dark energy \cite{Perrotta1999, Boisseau2000}. Indeed, scalar fields, originally introduced as inflatons driving the early inflationary epoch, have re-emerged as central ingredients in modern cosmology. However, they can also be employed to provide a connection between the early and late-time accelerated phases of the universe \cite{Poulin:2018cxd,Poulin:2023lkg,Odintsov:2016plw,Jaime:2022cho,Elizalde:2010ts,Chaussidon:2025npr,Carloni:2025jlk,DAgostino:2021vvv,Dunsby:2023qpb}.

Scalar field is essential for defining the physical theory within the scalar-tensor theories consistent with Mach’s principle. Although Mach’s principle inspired Einstein, General Relativity does not fully satisfy it. Brans-Dicke gravity \cite{Brans:1961sx}, is the first example of scalar-tensor theory, where the scalar field is coupled to the Lagrangian function of GR. Scalar-tensor theories, are defined in the so called Jordan frame, while GR is defined in the so-called Einstein frame. Under a conformal map, a mathematical relation that neutralizes the coupling function, any scalar-tensor theory can be expressed in the Einstein frame, that is, into an equivalent theory of GR with a scalar field minimally coupled to gravity. 
The conformal map relates solution trajectories into solution trajectories, for the gravitational field equations, but the physical properties of the solutions does not remain invariant under the conformal map. This has open a fundamental problem in modern cosmology, on which frame is physical \cite{Faraoni:1999hp}, for more details we refer the reader to \cite{Mukherjee:2023qan,Diaz:2023tma,Bamber:2022eoy,Karam:2017zno,Lobo:2016izs,Pandey:2016unk,Kamenshchik:2016gcy,Banerjee:2016lco,Bisabr:2012nq,Quiros:2011iv,Bhadra:2006rn,Alvarez:2001qj} and references therein.

Beyond Levi-Civita geometry, one can relax the assumptions of vanishing torsion and non-metricity. The metric teleparallel formulation, based on a torsionful but curvature-free connection, gives rise to $f(\mathbb{T})$ gravity \cite{Ferraro:2006jd,Ferraro:2011us}. A further alternative is the symmetric teleparallel approach, defined by non-metricity but zero curvature and torsion, which produces a parallel research direction towards $f(Q)$ gravity \cite{Jim2018,exactfq,heisen2024}.

Scalar fields non-minimally coupled to curvature or torsion have long been studied in cosmology \cite{Faraoni:2004pi,Poplawski2020,Skugoreva2015}. A natural analogue, the scalar-tensor non-metricity theory is the extension of the Symmetric Teleparallel General Relativity (STGR), in which the scalar field is incorporated into the physical space and interacts nontrivially with the fundamental Lagrangian of the theory, namely the non-metricity scalar $Q$ \cite{laur2024}. This has been developed as promising theoretical frameworks to address some of the most profound problems in modern cosmology, particularly the explanation of cosmic acceleration and the dynamics of the early universe. The widely studied $f(Q)$ gravity emerges as a particular case of scalar-non-metricity theory, corresponding to a specific choice of the scalar field configuration and its associated potential function \cite{Giacomini2024}. Considerable effort has been directed towards building unified models where a single scalar component behaves like dust in the early universe and drives acceleration at late times in this geometric trinity of scalar-tensor theories \cite{Basilakos2022,Saal2018, BD024, murtazascalar, Valcarcel2022}. 

In recent decades, scalar field cosmology has been enriched by the study of steep potentials, which can be written in the generic form
$$U(\phi) = U_0 e^{-\lambda \phi}, \qquad 
U(\phi) = U_0 e^{-\lambda \phi^n}, \quad n>1,$$
where the latter represents a steeper-than-exponential generalization. Such potentials naturally arise in higher-dimensional theories, string compactifications, and supergravity scenarios, and they have found widespread applications in cosmology. Exponential and steep potentials underpin many quintessence models of dark energy, ekpyrotic scenarios, and scaling solutions where the scalar sector tracks the background radiation or matter \cite{Copeland1998, Barreiro2000}. These forms are particularly appealing because they can alleviate the coincidence problem by dynamically adjusting the scalar contribution across different epochs. Moreover, double-exponential potentials have been explored to generate transitions between scaling regimes and late-time acceleration \cite{Ng2001, Guo2003}. In ekpyrotic cosmology, very steep negative exponential potentials yield stable anisotropy-suppressing attractors during contracting phases, providing alternatives to inflation \cite{Khoury2001, Lehners2008}. Thus, steep potentials serve as a unifying theme across early and late-time cosmology.

Copeland, Liddle, and Wands \cite{Copeland1998} demonstrated that for a scalar field with an exponential potential in GR, if the potential is steep, the field does not drive acceleration but instead enters a scaling regime, where its energy density tracks the background fluid (radiation or matter). These scaling solutions are the unique late-time attractors, and nucleosynthesis constraints rule out standard inflation with steep potentials due to the relic density problem. Shahalam et al. \cite{Yang2017}, showed that in scalar field dark energy models, steep exponential potentials can retain tracker behavior, where the field energy density scales with the background, and eventually exit this regime to drive late-time cosmic acceleration. In \cite{Das2019jcap}, the scalar field model with a steep potential was analyzed, and the authors claimed that higher values of the steepness index lead to unstable cosmological solutions. However, we do not agree with their conclusion, as the methodology adopted for applying the center manifold theory appears inconsistent. In \cite{Hossain2015}, quintessential inflation models based on canonical and noncanonical scalar fields with steeper-than-exponential potentials were analyzed, yielding successful inflation consistent with Planck 2015 data. Through tracker solutions and scalar–neutrino coupling, these models also recover the standard thermal history and naturally connect inflation to late-time cosmic acceleration.
Subsequent studies extended this framework to steeper or double-exponential forms, revealing richer phase-space structures and transitions between matter-like, ekpyrotic, and accelerating regimes \cite{Barreiro2000, Ng2001, Guo2003}. A few other interesting studies by considering the steep potential can be found in \cite{Lid2001,Sa2001}.
 
Dynamical system analysis (DSA) provides a powerful framework for analyzing models with steep potentials. By reformulating the field equations into autonomous systems of dimensionless variables, we are able to study the phase-space of the field equations by classifying critical points and study their stability. Each critical point describes an asymptotic solution for the field equations, attributed to a specific epoch of the cosmological history. Importantly, for steep or double-exponential potentials, DSA reveals non-trivial attractor behavior: the scalar field may act as a tracker, or, under certain couplings, drive acceleration despite the steepness of the potential \cite{Curbelo:2005dh,Lazkoz:2006pa,Lazkoz:2007mx,Leon:2008de,Billyard:2000bh,Coley:1999mj,Paliathanasis2014, Leon2009}. These results show that steep potentials, while often assumed too restrictive, can in fact yield a wide class of viable cosmological trajectories. Due to the importance of the DSA in the study of gravitational models, it has been widely applied in the literature \cite{Chakraborty:2024ybx,Paliathanasis:2024abl,vanderWesthuizen:2025vcb,Caldera-Cabral:2010yte,Carloni:2024ybx,Duchaniya:2024vvc,Kadam:2024fgz,Abramo:2003cq,Shah:2020puj,Chakraborty:2023clu,Paliathanasis:2023nkb,Murtaza:2025gme,Shabani:2025qxn,Khyllep:2022spx,Csillag:2025gnz,Rendall:2005fv}.

Motivated by these considerations, in this work, we focus on scalar-non-metricity gravity and perform a DSA treatment of a flat FLRW spacetime. This enables us to recast the field equations into an autonomous system, identify the fixed points, and characterize the physical nature of their associated asymptotic solutions, with particular emphasis on the role of steep scalar potentials in shaping the universe’s past and future evolution. To the best of our knowledge, this is the first systematic dynamical systems analysis of steep scalar potentials within scalar-non-metricity gravity, aiming to assess whether such models can reproduce viable cosmological epochs while offering novel departures from standard scalar-tensor scenarios. The structure of the paper is as follows.

 We give a concise overview of the mathematical framework of symmetric teleparallel theory, and then derive the corresponding field equations within the non-metricity based scalar-tensor gravity in Section \ref{secnew}. In Section \ref{sec1}, we investigate the cosmological aspects of this theory within the framework of compatible connection classes. The subsequent Sections ~\ref{sec4}, \ref{sec5}, and \ref{sec6}, present a comprehensive dynamical systems analysis, encompassing both finite and asymptotic regimes, for steep potential and power law coupling function. The analysis is performed across all three connection branches in the context of a spatially flat FLRW background. A brief comparison of our findings with some existing studies on non-minimally coupled scalar-curvature and scalar-torsion theories can be found in Section \ref{comparison}. Finally, our main findings are summarized in Section \ref{sec7}.

\section{Mathematical Framework of Scalar--nonmetricity Gravity}\label{secnew}

 In the context of Riemannian geometry, the torsion-free and metric-compatible affine connection is uniquely given by the Levi-Civita connection $\mathring{\Gamma}^\alpha{}_{\mu\nu}$. Its explicit form in terms of the metric $g_{\mu\nu}$ is given as
\begin{equation}
\mathring{\Gamma}^\alpha_{\,\,\,\mu\nu}=\tfrac{1}{2}g^{\alpha\beta}\left(\partial_\nu g_{\beta\mu}+\partial_\mu g_{\beta\nu}-\partial_\beta g_{\mu\nu}\right).
\end{equation}
To move beyond this framework, one may instead consider a torsionless and curvatureless connection $\Gamma^\alpha{}_{\mu\nu}$ that admits a non-vanishing non-metricity tensor, defined by
\begin{equation} \label{Q tensor}
Q_{\lambda\mu\nu} := \nabla_\lambda g_{\mu\nu}
= \partial_\lambda g_{\mu\nu}
- \Gamma^{\beta}_{\,\,\,\lambda\mu} g_{\beta\nu}
- \Gamma^{\beta}_{\,\,\,\lambda\nu} g_{\beta\mu} \neq 0 \,.
\end{equation}
This connection can be decomposed as
\begin{equation} \label{connc}
\Gamma^\lambda{}_{\mu\nu} = \mathring{\Gamma}^\lambda{}_{\mu\nu}+L^\lambda{}_{\mu\nu},
\end{equation}
where $L^\lambda{}_{\mu\nu}$ is the disformation tensor,
\begin{equation} \label{L}
L^\lambda{}_{\mu\nu} = \tfrac{1}{2}\left(Q^\lambda{}_{\mu\nu} - Q_\mu{}^\lambda{}_\nu - Q_\nu{}^\lambda{}_\mu \right).
\end{equation}

The tensor known as the superpotential, or equivalently the non-metricity conjugate, is expressed as
\begin{equation} \label{P}
P^\lambda{}_{\mu\nu} = \tfrac{1}{4} \left( -2 L^\lambda{}_{\mu\nu} + Q^\lambda g_{\mu\nu} - \tilde{Q}^\lambda g_{\mu\nu} - \delta^\lambda{}_{(\mu} Q_{\nu)} \right),
\end{equation}
with
\begin{equation*}
Q_\mu := Q_\mu{}^\nu{}_\nu = g^{\nu\lambda} Q_{\mu\nu\lambda}, 
\qquad 
\tilde{Q}_\mu := Q_{\nu\mu}{}^\nu = g^{\nu\lambda} Q_{\nu\mu\lambda}.
\end{equation*}
The non-metricity scalar $Q$ is then defined by
\begin{equation} \label{Q}
Q = Q_{\alpha\beta\gamma}P^{\alpha\beta\gamma}.
\end{equation}

It is useful to recall that Einstein’s General Relativity (GR) can be equivalently formulated in three distinct but dynamically equivalent ways, known as the geometric trinity of gravity.  
\begin{itemize}
    \item In the standard metric formulation, gravity is described by curvature via the Ricci scalar $\mathring{R}$. This leads to the Einstein--Hilbert action
    \begin{equation}
    S_{\rm EH} = \frac{1}{2\kappa}\int \sqrt{-g}\, \mathring{R}\, d^4x.
    \end{equation}

    \item In the metric teleparallel framework, the setup involves a connection with curvature and non-metricity set to zero, but with non-vanishing torsion. The torsion scalar is defined as $\mathbb{T} = S^\alpha{}_{\mu\nu} T_\alpha{}^{\mu\nu}$, and the teleparallel equivalent of GR (TEGR) is given by
    \begin{equation}
    S_{\rm TEGR} = \frac{1}{2\kappa}\int \sqrt{-g}\, \mathbb{T} \, d^4x.
    \end{equation}
    Generalizing this action yields $f(\mathbb{T})$ gravity, which has been extensively studied in cosmology.

    \item In the symmetric teleparallel framework, one assumes a connection that is torsion-free and curvature-free but endowed with non-zero non-metricity. The action of the symmetric teleparallel equivalent of GR (STEGR) is built from the non-metricity scalar $Q$, namely
    \begin{equation}
    S_{\rm STEGR} = \frac{1}{2\kappa}\int \sqrt{-g}\, Q \, d^4x,
    \end{equation}
    and its generalization leads to $f(Q)$ gravity.
\end{itemize}

Thus, scalar-torsion gravity theories extend the teleparallel framework by introducing a scalar field non-minimally coupled to $\mathbb{T}$, while scalar-non-metricity theories represent the analogous extension in the symmetric teleparallel setting, where the scalar couples to $Q$. Both frameworks allow for rich cosmological phenomenology and provide natural arenas to investigate dynamical alternatives to dark energy.

In analogy with the scalar-tensor extensions of GR, the scalar-non-metricity theory is described by the action \cite{Saal2018}
\begin{align}\label{eqn:ST}
S = \frac{1}{2\kappa}\int \sqrt{-g}\,\Big[ f(\phi)Q - h(\phi)\nabla^\alpha\phi\nabla_\alpha\phi - U(\phi) + 2\kappa \mathcal{L}_m \Big] \, d^4x ,
\end{align}
where $f(\phi)$ denotes the coupling of the scalar field $\phi$ to the non-metricity scalar $Q$, $h(\phi)$ controls the kinetic term, and $U(\phi)$ is the scalar potential.  

Special choices of the functions reproduce well-known models. For instance, setting $f(\phi)=\phi$ and $h(\phi)=\omega/\phi$ yields the Brans--Dicke theory, with $\omega$ the Brans--Dicke parameter. Moreover, the action is invariant under scalar-field redefinitions, allowing one to absorb one of the coupling functions into a constant without loss of generality. This fact allows us to consider $h(\phi)$ to be a constant in our present analysis. By taking $f(\phi)=\phi$, $h(\phi)=0$, and defining $\phi=f'(Q)$, one recovers the $f(Q)$ class of theories with potential $U(\phi)=f'(Q)Q-f(Q)$.

Varying the action with respect to the metric leads to the metric field equations
\begin{align}
\kappa T_{\mu\nu}
=&f\m G_{\mu\nu} +
2f'P^\lambda{}_{\mu\nu}\nabla_\lambda \phi
-h\nabla_\mu\Phi\nabla_\nu\phi
+\frac12hg_{\mu\nu}\nabla^\alpha\phi\nabla_\alpha\phi+\frac12Ug_{\mu\nu} \,,
\label{eqn:FE1}
\end{align}
where 
$\m G_{\mu\nu}$ represents the Einstein tensor associated to the Levi-Civita connection, and
$T_{\mu\nu}$ denotes the stress energy tensor given as 
\begin{align*}
T_{\mu\nu}=-\frac 2{\sqrt{-g}}\frac{\delta(\sqrt{-g}\mathcal L_M)}{\delta g^{\mu\nu}}\,,
\end{align*}
here
(~)' represents the derivative with respect to $\phi$. The second field equations, by varying the action with respect to the scalar field $\phi$ are given as
\begin{align}\label{eqn:FE2}
f'Q+h'\nabla^\alpha\phi\nabla_\alpha\phi+2h\m\nabla^\alpha\m\nabla_\alpha \phi-U'=0.
\end{align}
Besides the metric tensor and scalar field $\phi$, the theory involves the affine connection components as dynamical variables, resulting in the connection field equations
\begin{align}\label{eqn:FE3}
(\nabla_\mu-\tilde L_\mu)(\nabla_\nu-\tilde L_\nu)
\left[4fP^{\mu\nu }{}_\lambda+\kappa\Delta_\lambda{}^{\mu\nu}\right]=0\,,
\end{align}
where 
\begin{align}\Delta_\lambda{}^{\mu\nu}=-\frac2{\sqrt{-g}}\frac{\delta(\sqrt{-g}\mathcal L_M)}{\delta\Gamma^\lambda{}_{\mu\nu}}\,,
\end{align}
is the hypermomentum tensor \cite{hyper}. 

The relation for the effective stress-energy tensor $T^{\text{eff}}_{\mu\nu}$ is given by
\begin{align*}
    f\m G_{\mu\nu}=\kappa T^{\text{eff}}_{\mu\nu}\,,
\end{align*}
where
\begin{equation} \label{T^eff}
 T^{\text{eff}}_{\mu\nu} 
 =  T_{\mu\nu}+ \frac 1{\kappa}\left[
        -2f'P^\lambda{}_{\mu\nu}\nabla_\lambda \phi
        +h\nabla_\mu\phi\nabla_\nu\phi 
        -\frac12hg_{\mu\nu}\nabla^\alpha\phi\nabla_\alpha\phi
        -\frac12 Ug_{\mu\nu}\right]\,.
\end{equation}
In (\ref{T^eff}), the additional term can be interpreted as a fictitious dark energy component that can induce late-time cosmic acceleration via negative pressure.
\begin{align}
T^{\text{DE}}_{\mu\nu}= \frac 1{f}\left[
        -2f'P^\lambda{}_{\mu\nu}\nabla_\lambda \phi
        +h\nabla_\mu\phi\nabla_\nu\phi 
        -\frac12hg_{\mu\nu}\nabla^\alpha\phi\nabla_\alpha\phi
        -\frac12 Ug_{\mu\nu}\right]\,.
\end{align}

In this paper, the stress energy tensor is assumed to be a perfect fluid, which is expressed as
\begin{align}
T_{\mu\nu}=pg_{\mu\nu}+(p+\rho)u_\mu u_\nu.
\end{align}
with $p$, $\rho$, and $u^\mu$ representing
the pressure, energy density, and four velocity of the fluid, respectively.
In our present study, we consider the steep potential given by $U(\phi)=U_0 e^{\alpha(\frac{\phi}{M_p})^n}$ and a power law form of the coupling function $f(\phi)=\beta \phi^m$, and for simplicity, we assume $M_p=1$ for our analysis. 


\section{The cosmological aspects}\label{sec1}
Following the cosmological principle, the universe can be characterized by the spatially flat FLRW spacetime, which is homogeneous and isotropic on a large scale. 
The line element is given by 
\begin{equation}
ds^{2}=- dt^{2}+a^{2}\!\left(t\right)\!\left(dx^{2}+dy^{2}+dz^{2}\right),
\label{eq.06}
\end{equation}
where $a(t)$ is the scale factor. For a comoving observer with four–velocity $u^{\mu}$ ($u^{\mu}u_{\mu}=-1$), the Hubble expansion rate reads $H=\frac{\dot{a}}{a}$, $\dot{a}=\frac{da}{dt}$. In symmetric teleparallel gravity, the affine connection is torsionless and curvatureless but has nonvanishing nonmetricity. Its choice is therefore not fixed by the metric alone. Demanding that the connection be flat, symmetric, and compatible with FLRW symmetries singles out three inequivalent families of homogeneous and isotropic connections~\cite{FLRW/connection}. These families encode an extra (purely inertial) degree of freedom which we parametrize by a time function $\gamma(t)$. Operationally, the connection field equation plays the role of a constraint that fixes the allowed time dependence of $\gamma$ (or, equivalently, of the connection scalars $C_i$), and thus determines which dynamical branch is realized. This makes the choice of affine connection physically consequential: different connection classes lead to different expressions for $Q$ and different conditions for the existence and stability of cosmological critical points.

For the metric~\eqref{eq.06}, the nonvanishing components of the FLRW-compatible, flat, and symmetric affine connections can be written as
\begin{align}
\Gamma^{t}{}_{tt}=C_{1},\qquad \Gamma^{t}{}_{ii}=C_{2},\qquad \Gamma^{i}{}_{ti}=C_{3},
\end{align}
with the three classes given by
\begin{enumerate}\label{casesConn}
\item[(I)] $C_{1}=\gamma(t)$, \ $C_{2}=C_{3}=0$;
\item[(II)] $C_{1}=\dfrac{\dot{\gamma}(t)}{\gamma(t)}+\gamma(t)$, \ $C_{2}=0$, \ $C_{3}=\gamma(t)$, with $\gamma(t)\neq0$;
\item[(III)] $C_{1}=-\dfrac{\dot{\gamma}(t)}{\gamma(t)}$, \ $C_{2}=\gamma(t)$, \ $C_{3}=0$, with $\gamma(t)\neq0$.
\end{enumerate}
The corresponding background equations of motion are summarized in a generic form:
\begin{align}
\kappa p
=&f\left(-2\dot H-3H^2-\frac k{a^2}\right)
    +\frac12\dot f\left(3C_3+\frac{C_2}{a^2}-4H\right)
    -\frac12h\dot \phi^2    +\frac12U\,,\\
\kappa\rho
=&f\left(3H^2+3\dfrac k{a^2}\right)
    +\frac12\dot f\left(3C_3-3\frac{C_2}{a^2}\right)
     -\frac12h\dot \phi^2    -\frac12U\,,\\
0=&\partial_j\phi=\partial_j f
    ; \quad j=1,2,3\,.
\end{align}
And, the modified continuity relation is given by
\begin{align}
\kappa\dot{\rho}+3\kappa H(\rho+p)
=\frac32\left[
    \dot f\left(3C_3H-\frac{2\dot C_2+C_2H}{a^2}\right)
    +\ddot f\left(C_3-\frac{C_2}{a^2}\right)\right]\,,
\end{align}
where the right hand side vanishes due to the connection field equations in each classes. The generic non-metricity scalar 
\begin{align}
    Q
=&3\left(-2H^2+2\frac k{a^2}+3HC_3+\dot C_3+\frac1{a^2}(HC_2+\dot C_2)\right)\,,
\end{align}
produces for each class of connection the respective non-metricity scalar
\begin{itemize}
    \item $Q=-6H^{2}$
    \item $Q=-6H^{2}+9\gamma\,H+3\dot{\gamma}$
    \item $Q=-3\left[2H^{2}-\frac{\gamma}{a^{2}}\,H-\frac{\dot{\gamma}}{a^{2}}\right]$
\end{itemize}
In the following sections, we investigate the existence and stability of cosmological critical points for individual branches.

\section{Connection I}\label{sec4}
For the first connection class, the Friedmann equations are written as follows\footnote{The equations of motion derived from connection class I exhibit equivalence with those found in scalar-torsion gravity. Nevertheless, to the best of our knowledge, the dynamical implications of steeper scalar field potentials have not been systematically explored within the scalar-torsion framework. This section aims to address this gap in the literature.}

\begin{align}\label{peq_1}
\kappa p&=(-2\dot H-3H^2)f-2H\dot f+\frac U2-\frac{h_0\dot \phi^2}{2},
\end{align}
\begin{align}\label{rho_eq1}  
\kappa\rho&=3H^2f-\frac U2-\frac{h_0\dot \phi^2}{2}.
\end{align}
The scalar field Eq (\ref{eqn:FE2}) gives
\begin{align}\label{scalareq_1}
-6H^2f' -2h_0(\ddot\phi+3H\dot \phi)-U'=0.
\end{align}
The continuity relation is given as follows
\begin{align}\label{cq_1}
\kappa\dot{\rho}+3\kappa H(\rho+p)=0.
\end{align}
From Eq (\ref{rho_eq1}), we have
\begin{align}
\frac{\kappa \rho}{3fH^2}=1-\frac{h_0 \dot{\phi}^2}{6fH^2}-\frac{U}{6fH^2},
\end{align}
In the present analysis, we assume the pressureless dust era. Here, the $\overline{(.)}$ denotes differentiation with respect to the e-folding number $N=lna$ or $d/dN$. We define the dimensionless dependent variables in the context of H-normalization \cite{Copeland1998}
\begin{align}
x^2=\frac{\dot{\phi^2}}{6fH^2},~~y=\frac{U}{3fH^2}, ~~s=\frac{\kappa \rho}{3fH^2}, ~~ \lambda=\frac{U' \sqrt{f}}{U}, ~~ \zeta =\frac{f'}{\sqrt{f}},~~\Gamma=\frac{U'' U}{U'^2},~~\Delta=\frac{f''f}{f'^2},
\end{align}
It is evident that for $n=1$, the steep potential reduces to a simple exponential potential, which immediately gives $\Gamma=1$. Hence, we assume $n\neq1$. For $m=1$, gives a vanishing value of $\Delta$, i.e., $\Delta=0$. Hence, we also avoid $m=1$ for our analysis. The constraint equation is
\begin{align} \label{constraint_1}
 s=1-h_0x^2 -\frac{y}{2}   .
\end{align}
By taking $p=0$, we obtain the following from Eq (\ref{peq_1})
\begin{align} \label{dotH_1}
 \frac{\dot{H}}{H^2}=-\frac{3}{2}-\sqrt{6}x\zeta-\frac{3h_0x^2}{2}+\frac{3y}{4} .
\end{align}

It is important to mention that the variables $x$ and $y$ are not restricted and can take values throughout the space of real numbers. Indeed, they can take values at infinity. So, in order to study the dynamical evolution at infinity, we also define the compactified variables for thorough analysis. 
\subsection{Finite Critical Point Analysis}
By making use of (\ref{scalareq_1}), (\ref{cq_1}), and (\ref{dotH_1}), the general form of the dynamical system can be expressed as follows
\begin{align}
\overline{x}=\frac{\Big(-\left((\sqrt{6}\lambda+3h_0x)y\right)+2(-1+h_0x^2)(3h_0x+\sqrt{6}\zeta)\Big)}{4h_0},
\end{align}
\begin{align}
\overline{y}=\frac{y\Big(6+6h_0x^2-3y+2\sqrt{6}x(\lambda+\zeta)\Big)}{2},
\end{align}
\begin{align}
\overline{s}=-3s(1+\omega)+\sqrt{6}xs \zeta+3s+3h_0x^2s-\frac{3}{2}ys,
\end{align}
\begin{align}
 \overline{\lambda}=\sqrt{6}x\lambda^2(\Gamma-1)+\frac{\sqrt{3}x\lambda \zeta}{\sqrt{2}},
\end{align}
\begin{align}
 \overline{\zeta}=\sqrt{6}x\zeta^2(\Delta -\frac{1}{2}).
\end{align}
The equation of state (Eos) parameter for total fluid i.e $w_{eff}=-1-\frac{2}{3}\frac{\dot{H}}{H^2}$, is given as
\begin{align}
    w_{eff}=\frac{2\sqrt{2}x \zeta}{\sqrt{3}}+h_0x^2 -\frac{y}{2},
\end{align}
The deceleration parameter $q=-1-\frac{\dot{H}}{H^2}$, is given as
\begin{align}
   q=\frac{1}{2}+\sqrt{6}x\zeta +\frac{3h_0x^2}{2}-\frac{3y}{4}. 
\end{align}
We have 
\begin{align}\label{steepGamma}
    \Gamma=1+(n-1)(n\alpha)^{\frac{2-m}{2n-2+m}} \beta^{\frac{n}{2n-2+m}}\frac{1}{\lambda^{\frac{2n}{2n-2+m}}}, ~~\Delta=\frac{m-1}{m}.
\end{align}
Under these settings, and by employing the constraint in Eq (\ref{constraint_1}) to reduce the dimensionality of the phase space, the autonomous system can be reformulated as
\begin{align}\label{xeq_conl}
\overline{x}=\frac{\Big(-\left((\sqrt{6}\lambda+3h_0x)y\right)+2(-1+h_0x^2)(3h_0x+\sqrt{6}\zeta)\Big)}{4h_0},
\end{align}
\begin{align}
\overline{y}=\frac{y\Big(6+6h_0x^2-3y+2\sqrt{6}x(\lambda+\zeta)\Big)}{2},
\end{align}
\begin{align}
 \overline{\lambda}=\lambda\left(\sqrt{6}x(n-1)(n\alpha)^{\frac{2-m}{2n-2+m}}\beta^{\frac{n}{2n-2+m}}\lambda^{\frac{m-2}{2n-2+m}}+\frac{\sqrt{3}x \zeta}{\sqrt{2}}\right),
\end{align}
\begin{align}\label{zetaeq_conl}
 \overline{\zeta}=\frac{\sqrt{3}(m-2)x\zeta^2}{m\sqrt{2}}
\end{align}

\begin{table}[tbp]
    \centering
    \begin{tabular}{|c|c|c|c|c|}
        \hline
      \textbf{  Critical point}  &$(x,y,\lambda,\zeta)$ &\textbf{Existence }&\textbf{$w_{\text{eff}}$} & $q$ \\ \hline
       $A$ & $\left( 0,0,\lambda, 0 \right)$& $\lambda~$arbitrary & $0$&$\frac{1}{2}$ \\ \hline
         $B$ & $\left( 0,2,0, 0 \right)$ & Always&$-1$&$-1$\\ \hline
          $C$ & $\left( \frac{1}{\sqrt{h_0}},0,0,0 \right)$ & $h_0>0$&$1$&$2$\\ \hline
          \end{tabular}
     \caption{Critical points and their physical properties.}
    \label{table1_con1}
\end{table}

\begin{table}[tbp]
    \centering
     \begin{tabular}{|c|c|c|}
        \hline
       \textbf{ Critical point}  & \textbf{Eigenvalues} &\textbf{ Stability}  \\ \hline
        $A$ & $\left( 0,0,-\tfrac{3}{2}, 3 \right)$ & unstable \\ \hline
        $B$ & $\left(0,0,-3,-3\right)$
         & non-hyperbolic \\ \hline
        $C$ & 
         $\left(0,3,6,0 \right)$
         & unstable \\ \hline
    \end{tabular}
    \caption{Eigenvalues and Stability.}
    \label{table2_con1}
\end{table}
The properties of the critical points (CPs) corresponding to the dynamical system
(\ref{xeq_conl})-(\ref{zetaeq_conl}) are summarized in Table \ref{table1_con1} and \ref{table2_con1}, including their existence conditions, $w_{eff}$, $q$, eigenvalues and stability analysis. 

The critical point $A$  describes a decelerating dust-dominated universe. Its eigenvalues are $A$ are $e_{1}(A)=0, e_{2}(A)=0, e_{3}(A)=-\frac{3}{2}$ and $e_{4}(A)=3$, indicating its unstable nature. 

The critical point $B$ always exists, and its associated physical parameters are $w_{eff}=-1$ and $q=-1$, which correspond to a de Sitter solution. The eigenvalue at this point is given by $e_{1}(B)=0, e_{2}(B)=0$, $e_{4}=-3$ and $e_{4}(B)=-3$. This indicates that point $B$ is non-hyperbolic, and therefore its stability properties cannot be fully determined using linear stability theory alone. Since this point represents a potentially dominant de Sitter state, a careful stability analysis is particularly important. To this end, we apply the Center Manifold Theory (CMT), which allows us to identify specific parameter ranges under which the critical point $B$ behaves either as a late-time attractor or as an unstable solution. The detailed CMT analysis, presented in Appendix \ref{B_stability}, shows that for $m=2$, and $n>1$, the point exhibits instability, while for $m \geq 3$ and $n>1$ it describes a stable de Sitter attractor when $h_0>0$ and an unstable behavior for $h_0<0$.

 The physical parameters of the critical point $C$ are $w_{eff}=1$ and $q=2$, indicating that it corresponds to a decelerated stiff fluid universe. The eigenvalues corresponding to this point $e_{1}(C)=0, e_{2}(C)=3, e_{3}(C)=6$ and $e_{4}(C)=0$, which gives its unstable behavior.

\subsection{Analysis at infinity}
As discussed earlier, the variables are not constrained, and thus, to investigate the asymptotic behavior of the dynamical system (\ref{xeq_conl})-(\ref{zetaeq_conl}), we employ the Poincare compactification by introducing the corresponding compact variables as follows
\begin{align*}
    x=\frac{X}{\rho}, ~~y=\frac{Y}{\rho}.
\end{align*}
where $\rho= \sqrt{1-X^2-Y^2}$.  We define the new independent variable $dt=\sqrt{1-X^2-Y^2} dT$, where $X^2 \leq 1$ and $Y^2 \leq 1$ with constraint $1-X^2-Y^2 \geq0$. It is worth emphasizing that this approach provides a rigorous framework to investigate critical points at infinity and to assess whether Big Rip or Big Crunch singularities arise as asymptotic attractors, a feature that would ultimately undermine the physical viability of the theory.

At infinity, the dynamical system (\ref{xeq_conl})-(\ref{zetaeq_conl}) becomes\footnote{The symbolic notations $F_1,~F_2,~F_3$ and $F_4$ are used; the complete dynamical system equations at infinity are provided in Appendix \ref{conn1Infty}.}

\begin{equation}
\begin{aligned}
\frac{dX}{dT} =F_1(X,Y,\lambda,\zeta;m,n,h_0,\alpha,\beta),
\end{aligned}
\end{equation}
\begin{equation}
\begin{aligned}
\frac{dY}{dT} = 
F_2(X,Y,\lambda,\zeta;m,n,h_0,\alpha,\beta),
\end{aligned}
\end{equation}

\begin{align}
 \frac{d \lambda}{dT}=F_3(X,Y,\lambda,\zeta;m,n,h_0,\alpha,\beta),
\end{align}
\begin{align}
 \frac{d \zeta}{dT}=F_4(X,Y,\lambda,\zeta;m,n,h_0,\alpha,\beta).
\end{align}

The EoS and deceleration parameters in terms of these new variables are given as
\begin{align}
    w_{eff}=\frac{6h_0X^2+4\sqrt{6} X \zeta \rho-3Y\rho}{6\rho^2},
\end{align}
\begin{align}\label{Qef_conl}
   q=\frac{6h_0X^2+4\sqrt{6} X \zeta \rho-3Y\rho+2\rho^2}{4\rho^2}. 
\end{align}
\begin{table}[tbp]
    \centering
    \begin{tabular}{|c|c|c|c|c|c|}
        \hline
       \textbf{ Critical point}  &$(X,Y,\lambda,\zeta)$ &\textbf{Existence }&\textbf{$w_{\text{eff}}$} & $q$ &\textbf{ Stability}\\ \hline
       $A_{1\pm}$ & $\left( 0,\pm 1,0,0 \right)$ & Always&$\pm \infty$&$\pm \infty$& unstable\\ \hline
           $A_{2}$ & $\left( 1,0,0,0 \right)$ & Always&$\infty$&$\infty$& unstable\\ \hline
          \end{tabular}
     \caption{Critical points and their physical properties.}
    \label{table3_conl}
\end{table}

\begin{figure}[tbp]
    \centering
    \includegraphics[width=0.4\textwidth]{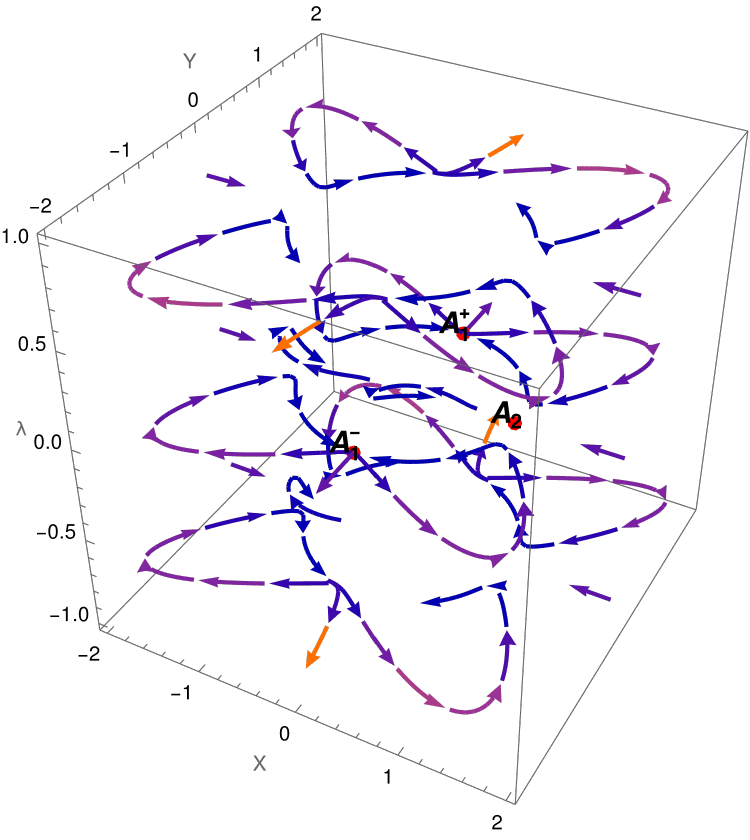}
    \caption{3D phase portrait for the dynamical system given in Eqs. (\ref{Xeq_conl})-(\ref{Zetaeq_conl}) for $h_0=0.5,~m=2,n=2,\alpha=1,\beta=2$.
   }
    \label{fig1_conl}
\end{figure}

\begin{figure}[tbp]
    \centering
    \begin{subfigure}[b]{0.45\textwidth}
        \centering
        \includegraphics[width=\textwidth]{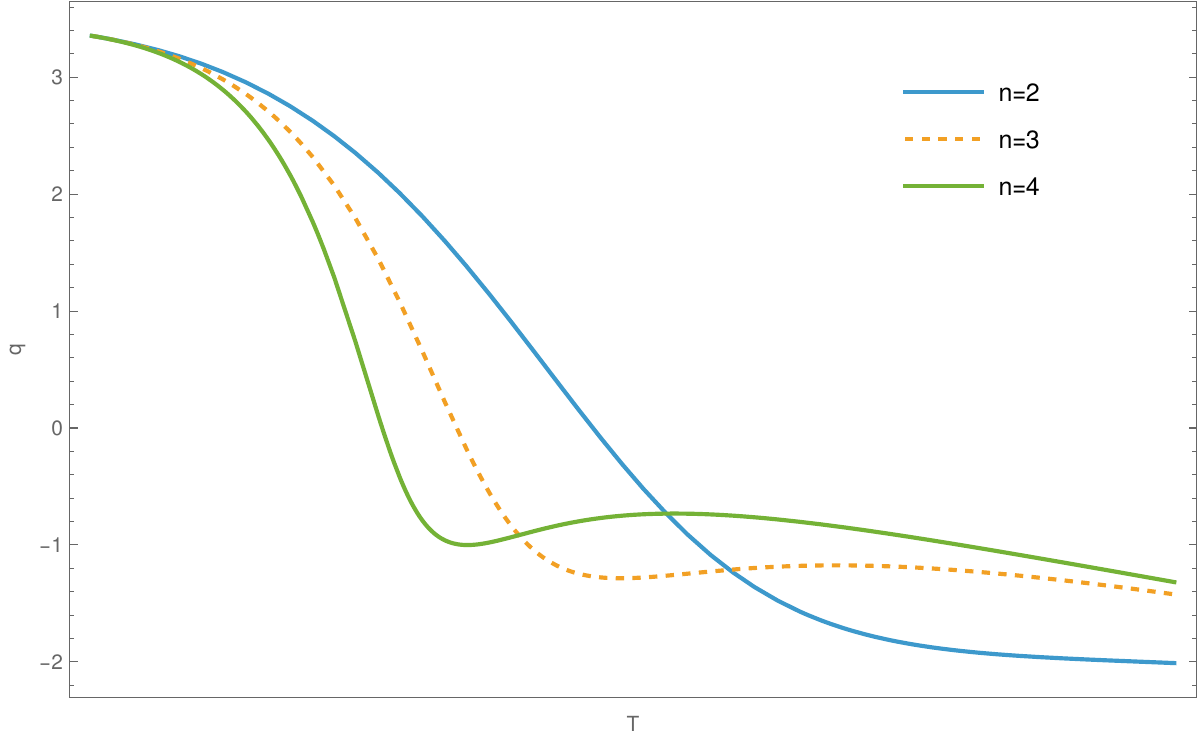}
        \label{fig:subweff1}
    \end{subfigure}
    \begin{subfigure}[b]{0.45\textwidth}
        \centering
        \includegraphics[width=\textwidth]{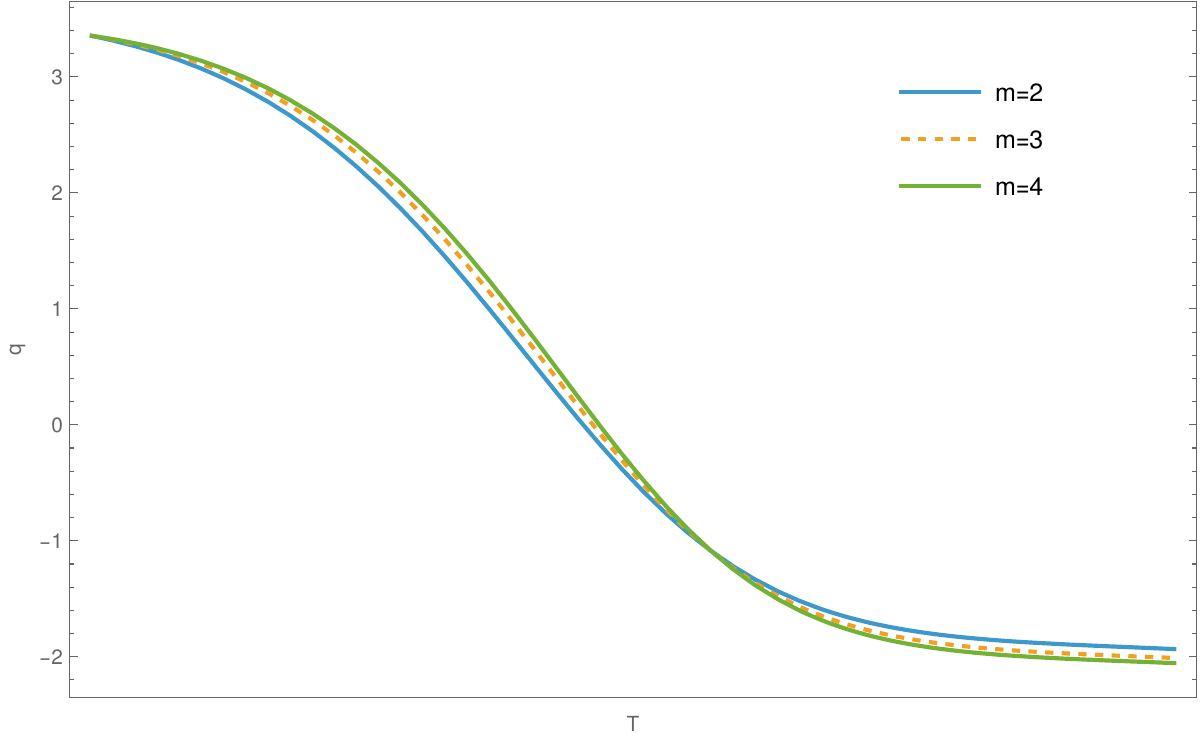}
        \label{fig:subweff2}
    \end{subfigure}
    
    \caption{Qualitative evolution of the deceleration parameter of the dynamical system (\ref{Xeq_conl})-(\ref{Zetaeq_conl})
for different values of $n$ and $m$, with initial conditions ($X[0]=0.8,~Y[0]=0.3 ,~\lambda[0]=0.9,~\zeta[0]=0.4$) with parameters values are $h_0=0.5,~\beta=2,~\alpha=1$.}
    \label{fig2_conl}
\end{figure}

The existence of critical points at infinity and their physical properties are summarized in Table \ref{table3_conl}. The analysis shows that the critical points $A_{1_\pm}$ and $A_{2}$ correspond to cosmological scenarios with $w_{eff}=\pm \infty$ and $q=\pm\infty$, representing either a Big Crunch or a Big Rip. Although the explicit stability analysis is omitted, we observe that the critical points at infinity, when present, invariably act as saddle points. The 3D phase portrait in Fig \ref{fig1_conl}, further substantiates this interpretation, clearly displaying their unstable nature. Moreover, Fig \ref{fig2_conl} exhibits the qualitative evolution of the deceleration parameter (\ref{Qef_conl}), for different values of $m$ and $n$.


\section{Connection II}\label{sec5}
The Friedmann equations corresponding to the connection class II are expressed as follows
\begin{align}\label{peq_2}
\kappa p
=&\left(-2\dot H-3H^2\right)f
    +\frac12\left(3\gamma-4H\right)\dot f
    -\frac12h_0\dot \phi^2    +\frac12U\,,
\end{align}
\begin{align}\label{rhoeq}
\kappa\rho
=&3H^2f
    +\frac32\gamma\dot f
     -\frac12h_0\dot \phi^2    -\frac12U\,.
\end{align}
For this connection from Eq (\ref{eqn:FE2}) which is a scalar field equation, we have
\begin{align}\label{eqn:FE2-2}
\left(-6H^2+3\left\{\dot\gamma+3H\gamma\right\}\right)f'
 -2h_0(\ddot\phi+3H\dot \phi)-U'=0.
\end{align}
The corresponding connection equation is written as
\begin{align} \label{connection:eq}
3\gamma \left(3\dot f H +\ddot f\right)=0.
\end{align}
Below, the continuity relation is provided, in which the right-hand side identically vanishes as a consequence of (\ref{connection:eq})
\begin{align}
\kappa\dot{\rho}+3\kappa H(\rho+p)
=\frac32\gamma\left[  3\dot f H +\ddot f \right].
\end{align}
We observe that the additional degree of freedom contributes to the system solely through terms involving derivatives of the coupling function  $f(\phi)$, and therefore becomes relevant only in the presence of a non-minimal coupling.
We can write the Eq (\ref{rhoeq}) as
\begin{align}
\frac{\kappa \rho}{3fH^2}=1+\frac{\gamma \dot{\phi}f'}{2fH^2}-\frac{h_0 \dot{\phi}^2}{6fH^2}-\frac{U}{6fH^2},
\end{align}
Definition of dimensionless variables within the H-normalization framework
\begin{align}\label{connIIvar}
x^2&=\frac{\dot{\phi^2}}{6fH^2},~~y=\frac{U}{3fH^2},~~z=\frac{\sqrt{3}\gamma f'}{\sqrt{2f}H}, ~~s=\frac{\kappa \rho}{3fH^2},
\notag\\&\lambda=\frac{U' \sqrt{f}}{U}, ~~ \zeta =\frac{f'}{\sqrt{f}},~~\Gamma=\frac{U'' U}{U'^2},~~\Delta=\frac{f''f}{f'^2}.
\end{align}
The constraint equation is
\begin{align} {\label{constr}}
 s=1+xz-h_0x^2 -\frac{y}{2}   .
\end{align}
From Eq (\ref{peq_2}) by considering $p=0$, we can get following
\begin{align} {\label{dH_2}}
 \frac{\dot{H}}{H^2}=-\frac{3}{2}+\frac{3xz}{2}-\sqrt{6}x\zeta-\frac{3h_0x^2}{2}+\frac{3y}{4}   ,
\end{align}
Likewise, variables $x$, $y$, and $z$ are not subject to any upper bounds and may take values across the entire real space. In particular, their dynamical trajectories can extend to infinity. To properly investigate the behavior of the system in this asymptotic regime, we also introduce compactified variables, which allow for a comprehensive analysis of the evolution, including at infinity. 
\subsection{Finite Critical Point Analysis}
 The general dynamical system equations by utilizing the relations (\ref{eqn:FE2-2}), (\ref{connection:eq}) and (\ref{dH_2}) can be formulated as,
\begin{equation}
\overline{x}=-\frac{3x}{2}-\frac{3x^2 z}{2}+\frac{3h_0x^3}{2}-\frac{3xy}{4}- \sqrt{6}x^2 \Delta \zeta+\frac{x^2\sqrt{6}\zeta}{2},
\end{equation}
\begin{equation}
\overline{y}=\sqrt{6}\lambda xy+3y-3yzx+3h_0x^2 y-\frac{3y^2}{2}+\sqrt{6}xy\zeta,
\end{equation}
\begin{equation}
    \overline{z}=\sqrt{6}\zeta-\frac{3z}{2}-2\sqrt{6}x^2h_0\Delta \zeta+\frac{\sqrt{3}\lambda y}{\sqrt{2}}+\sqrt{6}xz \Delta \zeta-\frac{3z^2 x}{2}+\frac{\sqrt{3}xz\zeta}{\sqrt{2}}+\frac{3h_0x^2z}{2}-\frac{3yz}{4},
\end{equation}
\begin{align}
\overline{s}=-3s(1+\omega)+\sqrt{6}xs \zeta-3sxz+3s+3h_0x^2s-\frac{3}{2}ys,
\end{align}
\begin{align}
 \overline{\lambda}=\sqrt{6}x\lambda^2(\Gamma-1)+\frac{\sqrt{3}x\lambda \zeta}{\sqrt{2}},
\end{align}
\begin{align}
 \overline{\zeta}=\sqrt{6}x\zeta^2(\Delta -\frac{1}{2}).
\end{align}
The EoS $w_{eff}$ and deceleration parameter $q$ are expressed as
\begin{align}
    w_{eff}=-xz+\frac{2\sqrt{2}x \zeta}{\sqrt{3}}+h_0x^2 -\frac{y}{2},
\end{align}
\begin{align}
   q=\frac{1}{2}-\frac{3xz}{2}+\sqrt{6}x\zeta +\frac{3h_0x^2}{2}-\frac{3y}{4}. 
\end{align}

Now, using (\ref{steepGamma}) and also the constraint Eq (\ref{constr}), we can rewrite the autonomous system as,

\begin{equation}\label{x_conll}
\overline{x}=-\frac{3x}{2}-\frac{3x^2 z}{2}+\frac{3h_0x^3}{2}-\frac{3xy}{4}- \frac{\sqrt{6}x^2\zeta}{2}+\frac{x^2\sqrt{6}\zeta}{m},
\end{equation}
\begin{equation}
\overline{y}=\sqrt{6}\lambda xy+3y-3yzx+3h_0x^2 y-\frac{3y^2}{2}+\sqrt{6}xy\zeta,
\end{equation}
\begin{equation}
    \overline{z}=\sqrt{6}\zeta-\frac{3z}{2}-\frac{2(m-1)\sqrt{6}x^2h_0\zeta}{m}+\frac{\sqrt{3}\lambda y}{\sqrt{2}}+\frac{\sqrt{6}(m-1)xz \zeta}{m}-\frac{3z^2 x}{2}+\frac{\sqrt{3}xz\zeta}{\sqrt{2}}+\frac{3h_0x^2z}{2}-\frac{3yz}{4},
\end{equation}
\begin{align}
 \overline{\lambda}=\lambda\left(\sqrt{6}x(n-1)(n\alpha)^{\frac{2-m}{2n-2+m}}\beta^{\frac{n}{2n-2+m}}\lambda^{\frac{m-2}{2n-2+m}}+\frac{\sqrt{3}x \zeta}{\sqrt{2}}\right),
\end{align}
\begin{align}\label{zeta_conll}
 \overline{\zeta}=\frac{\sqrt{3}(m-2)x\zeta^2}{m\sqrt{2}}.
\end{align}

\begin{table}[tbp]
    \centering
      \begin{tabular}{|c|c|c|c|c|}
        \hline
    \textbf{Critical point}  &$(x,y,z,\lambda,\zeta)$ &\textbf{Existence} &\textbf{$w_{\text{eff}}$} & $q$ \\ \hline
       $P_1$ & $\left( 0,0,z,\lambda, \frac{1}{2}\sqrt{\frac{3}{2}}z \right)$& $\lambda,z=$arbitrary & $0$&$\frac{1}{2}$ \\ \hline
         $P_2$ & $\left( 0,2,z_c,\frac{1}{2}(\sqrt{6}z_c-2\zeta_c), \zeta_c \right)$ &  $\zeta_c,z_c=$arbitrary&$-1$&$-1$\\ \hline
          $P_3$ & $\left(x,0,\frac{-1+h_0x^2}{x},0,0 \right)$ & $x \neq 0$&$1$&$2$\\ \hline
          \end{tabular}
    \caption{Critical points and their physical properties.}
    \label{table1_conll}
\end{table}

\begin{table}[tbp]
    \centering
    \begin{tabular}{|c|c|c|}
        \hline
        \textbf{Critical point}  & \textbf{Eigenvalues} & \textbf{Stability} \\ \hline
        $P_1$ & $\left( 0,0,-\tfrac{3}{2},-\tfrac{3}{2}, 3 \right)$ & unstable \\ \hline
        $P_2$ & $\left( 0,0,-3,-3,-3 \right)$ & non-hyperbolic \\ \hline
        $P_3$ & $\left( 0,0,3,6,0 \right)$ & unstable \\ \hline
    \end{tabular}
    \caption{Eigenvalues and Stability.}
    \label{table2_conll}
\end{table}
The detailed examination of the critical points, which specifies the existence conditions, stability nature, and cosmological parameters for the dynamical system (\ref{x_conll})-(\ref{zeta_conll}) is summarized in Table \ref{table1_conll} and \ref{table2_conll}.

The critical point $P_1$ represents a matter dominated universe, as $w_{eff}=0$, $ q=\frac{1}{2}$. The eigenvalues are $e_{1}(P_1)=0, e_{2}(P_1)=0, e_{3}(P_1)=-\frac{3}{2}$, $e_{4}(P_1)=-\frac{3}{2}$, and $e_{5}(P_1)=3$, giving its unstable behavior. 

The critical point $P_2$ describes a de Sitter solution as $w_{eff}=-1$ and $ q=-1$. It is non-hyperbolic point as the eigenvalues at this are $e_{1}(P_2)=0, e_{2}(P_2)=0$, $e_{3}(P_2)=-3,~e_{4}(P_2)=-3$, and $e_{5}(P_2)=-3$ meaning that linear stability fails to explain it stability behavior. Hence, we apply the center manifold theory to analyze its stability nature. The detail application of CMT for this point is given in the Appendix \ref{P2_stability}. It is observed that for $m=2$ and $n>1$, the point $P_2$ gives unstable behavior, while for $m \geq3$ and $n>1$, it has a stable nature.

The CP $P_3$ describes the decelerated universe of a stiff fluid since $w_{eff}=1$ and $q=2$. The point $P_3$ has an unstable nature confirmed by eigenvalues that are $e_{1}(P_3)=0, \,e_{2}(P_3)=0$, $e_{3}(P_3)=3,~e_{4}(P_3)=6$ and $e_{5}(P_3)=0$.
\subsection{Analysis at infinity}
In order to examine the existence of critical points at infinity, because our variables are not constrained, we introduce Poincare variables, which can be given as follows
\begin{equation*}
    x=\frac{X}{\rho}, ~~y=\frac{Y}{\rho},~~z=\frac{Z}{\rho}.
\end{equation*}
where $\rho= \sqrt{1-X^2-Y^2-Z^2}$. We define the new independent variable \\
$dt=\sqrt{1-X^2-Y^2-Z^2} dT$,  where $X^2 \leq 1$, $Y^2 \leq 1$, and $Z^2 \leq 1$ with constraint $1-X^2-Y^2-Z^2 \geq0$.  The dynamical system (\ref{x_conll})-(\ref{zeta_conll}) at infinity becomes\footnote{The symbolic notations $S_1,~S_2,~S_3,~S_4$ and $S_5$ are used; detailed calculations are provided in Appendix \ref{conn2Infty}.}
\begin{equation}
\begin{aligned}
\frac{dX}{dT}=
S_1(X,Y,Z,\lambda,\zeta;m,n,h_0,\alpha,\beta),
\end{aligned}
\end{equation}
\begin{equation}
\begin{aligned}
\frac{dY}{dT}=S_2(X,Y,Z,\lambda,\zeta;m,n,h_0,\alpha,\beta),
\end{aligned}
\end{equation}
\begin{equation}
\begin{aligned}
\frac{dZ}{dT}=S_3(X,Y,Z,\lambda,\zeta;m,n,h_0,\alpha,\beta),
\end{aligned}
\end{equation}
\begin{align}
 \frac{d \lambda}{dT}=S_4(X,Y,Z,\lambda,\zeta;m,n,h_0,\alpha,\beta),
\end{align}
\begin{align}
 \frac{d \zeta}{dT} =S_5(X,Y,Z,\lambda,\zeta;m,n,h_0,\alpha,\beta).
\end{align}

The equation of state parameter in terms of these new variables is expressed as
\begin{align}
    w_{eff}=\frac{6h_0X^2-6XZ+4\sqrt{6} X \zeta \rho-3Y\rho}{6\rho^2},
\end{align}
Also, the deceleration parameter is given as
\begin{align}\label{Wef_conll}
   q=\frac{6h_0X^2-6XZ+4\sqrt{6} X \zeta \rho-3Y\rho+2\rho^2}{4\rho^2}. 
\end{align}

\begin{table}[tbp]
\centering
\renewcommand{\arraystretch}{1.2}
\setlength{\tabcolsep}{4pt}    

\begin{adjustbox}{max width=\textwidth}
\begin{tabular}{|c|c|c|c|c|c|}
\hline
\textbf{Critical point} & \textbf{$(X,Y,Z,\lambda,\zeta)$} & \textbf{Existence} & \textbf{$w_{\text{eff}}$} & \textbf{$q$} & \textbf{Stability} \\ \hline
$B_{1\pm}$ & $\left(0,\pm \sqrt{1-Z^2},Z,0,0\right)$ & $-1 \le Z \le 1$ & $\pm \infty$ & $\pm \infty$ & unstable \\ \hline
$B_{2\pm}$ & $\left(0,0,\pm 1,0,0\right)$ & Always & $0$ & $\tfrac{1}{2}$ & unstable \\ \hline
$B_{3}$ & $\left(\sqrt{1-Z^2},0,Z,0,0\right)$ & $-1 \le Z \le 1$ & $\infty$ & $\infty$ & unstable \\ \hline
$B_{4\pm}$ & $\left(\tfrac{Z}{h_0},\pm \tfrac{\sqrt{h_0^2-Z^2-h_0^2Z^2}}{h_0},Z,0,0\right)$ & $h_0 \ne 0$,\; $-\sqrt{\tfrac{h_0^2}{1+h_0^2}} \le Z \le \sqrt{\tfrac{h_0^2}{1+h_0^2}}$ & $\pm \infty$ & $\pm \infty$ & unstable \\ \hline
\end{tabular}
\end{adjustbox}

\caption{Critical points and their physical properties.}
\label{table6}
\end{table}
\begin{figure}[tbp]
    \centering
    \begin{subfigure}{0.45\textwidth}
        \centering
        \includegraphics[width=\linewidth]{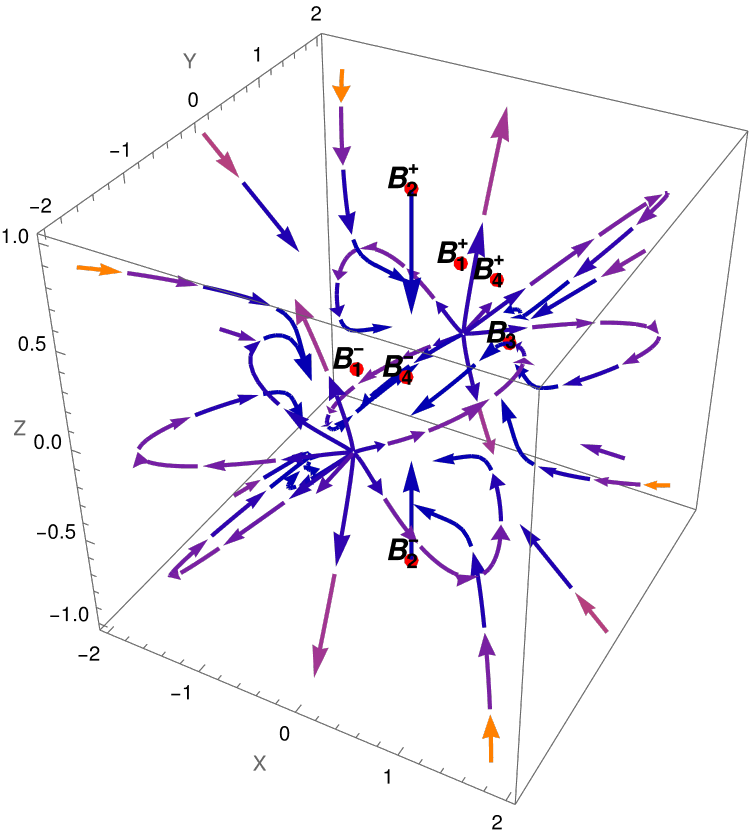}
        \caption{3D phase portrait for $h_0=0.5,~m=2,~n=2$.}
    \end{subfigure}
    \hfill
    \begin{subfigure}{0.45\textwidth}
        \centering
        \includegraphics[width=\linewidth]{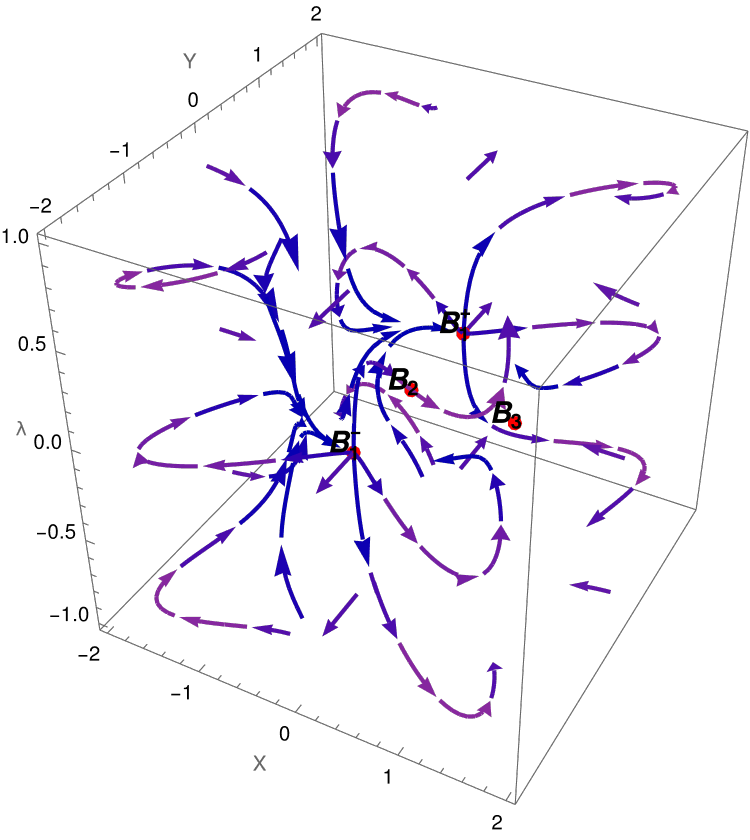}
        \caption{3D phase portrait for $h_0=0.5,~m=2,~n=2,~\beta=2,~\alpha=1$.}
    \end{subfigure}

    \caption{3D phase portraits for the dynamical system given in Eqs.~(\ref{Xeq_conll})-(\ref{Zetaeq_conll}). }
    \label{fig1_conll}
\end{figure}

\begin{figure}[tbp]
    \centering
    \begin{subfigure}[b]{0.45\textwidth}
        \centering
        \includegraphics[width=\textwidth]{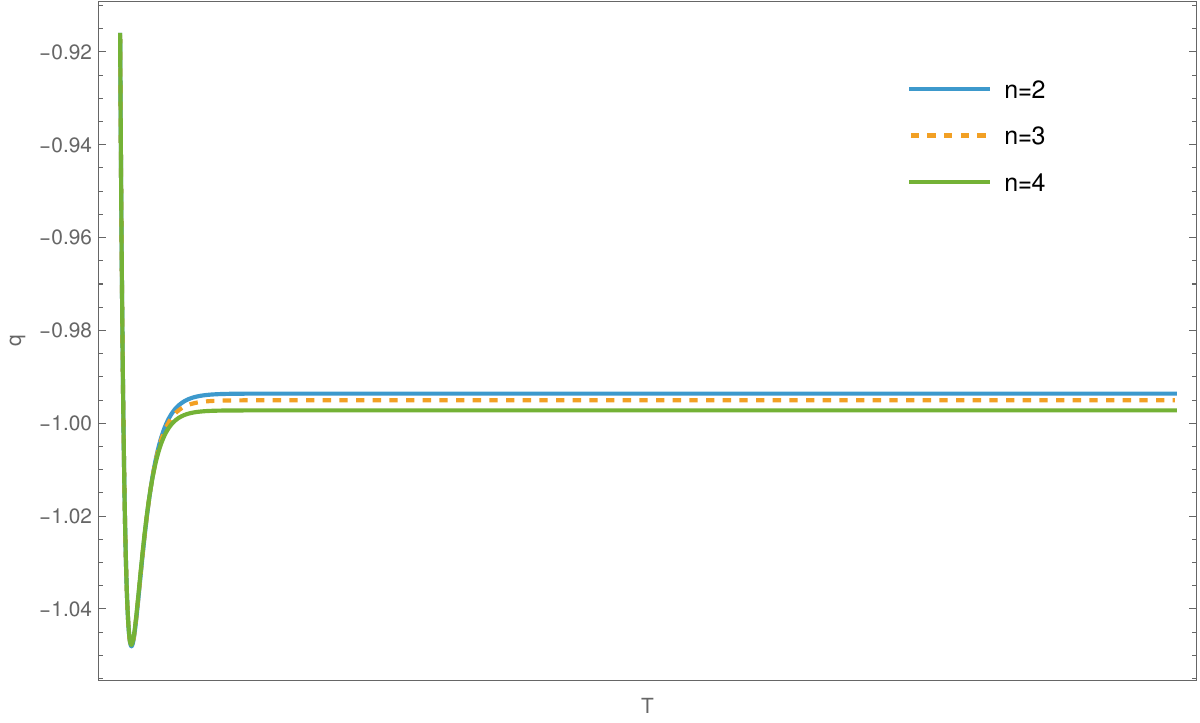}
    \end{subfigure}
    \begin{subfigure}[b]{0.45\textwidth}
        \centering
        \includegraphics[width=\textwidth]{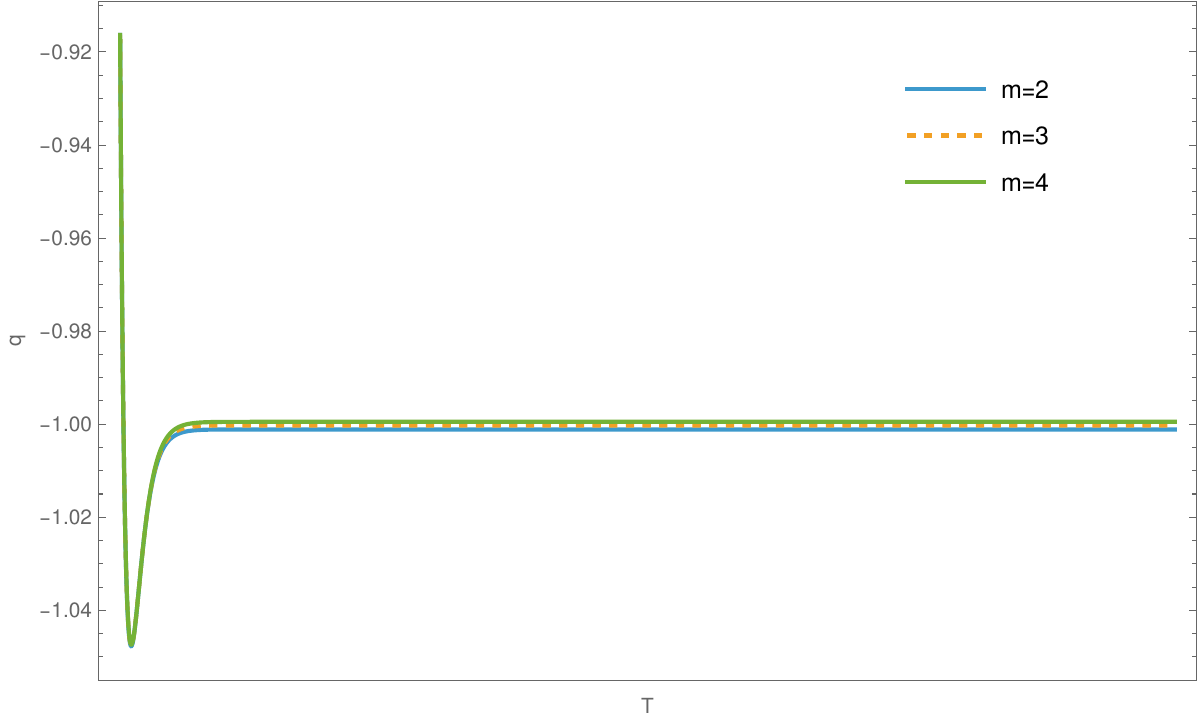}
    \end{subfigure}
    
    \caption{Qualitative evolution of the deceleration parameter of the dynamical system (\ref{Xeq_conll})-(\ref{Zetaeq_conll})
for different values of $n$ and $m$, with initial conditions ($X[0]=0.2,~Y[0]=0.7,~Z[0]=0.1 ,~\lambda[0]=0.9,~\zeta[0]=0.1$) with parameters values are $h_0=0.5,~\beta=2,~\alpha=1$.}
    \label{fig2_conll}
\end{figure}
The dynamical analysis at infinity reveals that the critical points $B_{1\pm}$,  $B_{3}$, and $B_{4\pm}$ correspond to extreme singularities such as a Big Rip or Big Crunch characterized by $w_{eff}=\pm \infty$ and $q=\pm \infty$ as mentioned in Table \ref{table6}. Moreover, the points $B_{2\pm}$ describe a matter dominated decelerated universe, with $w_{eff}=0$ and $q=\frac{1}{2}$. The 3D phase space portraits in Fig \ref{fig1_conll} clearly demonstrate the intrinsic instability of these critical points. The qualitative evolution of the deceleration parameter (\ref{Wef_conll}) is demonstrated in Fig \ref{fig2_conll} for the system (\ref{Xeq_conll})-(\ref{Zetaeq_conll}).

\section{Connection III}\label{sec6}
By considering $\frac{\gamma}{a^2}=\mathring{\gamma}$, the Friedmann equations  are given as

\begin{align}\label{peq}
\kappa p
=\; & f\left(-2\dot H-3H^2\right)
    +\frac12\dot f\left(\mathring{\gamma}-4H\right)
    -\frac12h_0\dot \phi^2    +\frac12U\,,\\
    \kappa\rho
=\; & 3H^2f
    -\frac32\dot f\mathring{\gamma}
     -\frac12h_0\dot \phi^2    -\frac12U\,.    
    \end{align}
The scalar field equation (\ref{eqn:FE2}) yields
\begin{align}\label{fieldeq}
\left(-6H^2
+3(\mathring{\dot\gamma}+3H\mathring{\gamma})\right)f'
 -2h_0(\ddot\phi+3H\dot \phi)-U'=0.
\end{align}
The connection field equation (\ref{eqn:FE3}) gives us
\begin{align} \label{conneq}
0
=-\frac32\left[
    \dot f\left(2\mathring{\dot \gamma}+5H\mathring{\gamma}\right)
    +\ddot f\mathring{\gamma}\right].
\end{align}
In the context of H-normalization, dimensionless dependent variables can be given as 
\begin{align}
\label{variables}
    x &= \frac{\dot{\phi}}{\sqrt{6f}H},
    y = \frac{U}{3H^2f}, 
   z = \frac{\mathring{\gamma}}{H},
    s = \frac{\kappa\rho}{3H^2f}, \notag \\
    \lambda &= \frac{U' \sqrt{f}}{U}, 
    \zeta = \frac{f'}{\sqrt{f}}, 
   \Gamma  = \frac{U'' U}{U'^2}, 
   \Delta = \frac{f''f}{f'^2}.
\end{align}
The relation for the constraint equation is given as
\begin{align}\label{const_3}
    s=1-\frac{\sqrt{3}}{\sqrt{2}}x \zeta z-h_0x^2-\frac{y}{2}.
\end{align}
Assuming $p=0$, from Eq (\ref{peq}) one can get the following
\begin{align} {\label{dH}}
\frac{\dot{H}}{H^2}=
- \frac{3}{2} - \frac{3}{2} h_0 x^2 + \frac{3 y}{4} - \sqrt{6} \, x \, \zeta + \frac{1}{2}\sqrt{\frac{3}{2}}x z \zeta  .
\end{align}
Again, the variables $x$, $y$, and $z$ are not constraints which means they can take values at infinity. To systematically explore the global dynamics, including the asymptotic regime, we also employ compactified variables, which provide a rigorous framework for analyzing the system’s evolution at infinity. 
\subsection{Finite Critical Point Analysis}
 The dynamical system equations by utilizing the equation (\ref{peq}), (\ref{fieldeq}), and (\ref{conneq}) is given as,
\begin{align}
\resizebox{\textwidth}{!}{$
\bar{x} = \frac{3x\Big(8\sqrt{6}h_0^2x^3-8 \lambda y-2h_0x^2(-8+z)\zeta+(-16+(10-3y)z)\zeta-\sqrt{6}x\left(4h_0(2+y)+z(-2+z+4\Delta)\zeta^2\right)\Big)}{4 \left(4 \sqrt{6} h_0 x  + 3 \zeta z\right)},
$}
\end{align}
\begin{align}
\resizebox{\textwidth}{!}{$
\bar{z} = \frac{z \Big(24\sqrt{6}h_0^2x^3+12\lambda y-3\left(-8+3(2+y)z\right)\zeta-6h_0x^2\left(z+8(-2+\Delta)\right)\zeta+\sqrt{6}x\left(8h_0-12h_0 y-3(-4+z)z\zeta^2)\right)\Big)}{ 4\left(4 \sqrt{6} h_0 x  + 3 \zeta z\right)},
$}
\end{align}
\begin{align}
\bar{y}=\frac{
y \left(
6+6 h_0 x^2 - 3 y  + 
\sqrt{6} x 
  \left(2\lambda-(-2+z\zeta)\right)
\right)
}{2 },
\end{align}
\begin{align}
\overline{s}=-3s(1+\omega)+\sqrt{6}xs \zeta-\sqrt{\frac{3}{2}}sxz\zeta+3s+3h_0x^2s-\frac{3}{2}ys,
\end{align}
\begin{align}
 \overline{\lambda}=\sqrt{\frac{3}{2}}\, \lambda x \left(
2 \lambda (-1 + \Gamma) + \zeta
\right),
\end{align}
\begin{align}
 \overline{\zeta}=\sqrt{\frac{3}{2}}\, x \left(-1 + 2 \Delta  \right) \zeta^2.
\end{align}
The physical parameters, such as $q$ and $w_{eff}$ are expressed in terms of variables as
\begin{align}
q=\frac{
   2 +6h_0x^2-3y-\sqrt{6}x(-4+z)\zeta
}{
    4
},
\end{align}
\begin{align}
w_{eff}=h_0 x^2 
    -\frac{y}{2} -
    \frac{x\zeta(-4+z)}{\sqrt{6}}.
\end{align}
Now using (\ref{steepGamma}) and (\ref{const_3}), the reduced autonomous dynamical system can be rewritten as,
\begin{align}\label{xeq_3}
\resizebox{\textwidth}{!}{$
\bar{x} =\frac{3x\Big(8\sqrt{6}h_0^2x^3-8 \lambda y-2h_0x^2(-8+z)\zeta+(-16+(10-3y)z)\zeta-\sqrt{6}x\left(4h_0(2+y)+z(2+z-\frac{4}{m})\zeta^2\right)\Big)}{4 \left(4 \sqrt{6} h_0 x  + 3 \zeta z\right)},
$}
\end{align}
\begin{align}
\resizebox{\textwidth}{!}{$
\bar{z} = \frac{z \Big(24\sqrt{6}h_0^2x^3+12\lambda y-3\left(-8+3(2+y)z\right)\zeta-6h_0x^2\left(z-8(1+\frac{1}{m})\right)\zeta+\sqrt{6}x\left(8h_0-12h_0 y-3(-4+z)z\zeta^2)\right)\Big)}{ 4\left(4 \sqrt{6} h_0 x  + 3 \zeta z\right)},
$}
\end{align}
\begin{align}
\bar{y}=\frac{
y \left(
6+6 h_0 x^2 - 3 y  + 
\sqrt{6} x 
  \left(2\lambda-(-2+z)\zeta\right)
\right)
}{
2 
},
\end{align}
\begin{align}
 \overline{\lambda}=\lambda\left(\sqrt{6}x(n-1)(n\alpha)^{\frac{2-m}{2n-2+m}}\beta^{\frac{n}{2n-2+m}}\lambda^{\frac{m-2}{2n-2+m}}+\frac{\sqrt{3}x \zeta}{\sqrt{2}}\right),
\end{align}
\begin{align}\label{zetaeq_3}
 \overline{\zeta}=\frac{\sqrt{3}(m-2)x\zeta^2}{m\sqrt{2}}.
\end{align}

\begin{table}[tbp]
    \centering
    \begin{tabular}{|c|c|c|c|c|}
        \hline
        \textbf{Critical point}  &$(x,y,z,\lambda,\zeta)$ &\textbf{Existence} &\textbf{$w_{\text{eff}}$} & $q$ \\ \hline
       $P$ & $\left( 0,0,\frac{4}{3},\lambda, \zeta \right)$& $\lambda,\zeta=$arbitrary & $0$&$\frac{1}{2}$ \\ \hline
         $Q$ & $\left( 0,2,z_c,\frac{1}{2}(3z_c-2)\zeta_c, \zeta_c \right)$ &  $\zeta_c,z_c=$arbitrary&$-1$&$-1$\\ \hline
         $R$ & $\left( \frac{1}{\sqrt{h_0}},0,0,0,0\right)$ &  $h_0>0$&$1$&$2$\\ \hline
          \end{tabular}
 \caption{Critical points and their physical properties.}
    \label{table1_conlll}
\end{table}
\begin{table}[tbp]
    \centering
    \begin{tabular}{|c|c|c|}
        \hline
       \textbf{ Critical point}  & \textbf{Eigenvalues} & \textbf{Stability} \\ \hline
        $P$ & $\left( 0,0,-\tfrac{3}{2},-\tfrac{1}{2}, 3 \right)$ & unstable \\ \hline
        $Q$ & $\left( 0,0,-5,-3,-3 \right)$ & non-hyperbolic \\ \hline
        $R$ & $\left( 0,2,3,6,0 \right)$ & unstable \\ \hline
        \end{tabular}
    \caption{Eigenvalues and Stability.}
    \label{table2_conlll}
\end{table}
The critical points are listed in Table \ref{table1_conlll} and \ref{table2_conlll}, which give their existence conditions, cosmological parameters $w_{eff}$, $q$, and attractor nature. 

The critical point $P$ has $w_{eff}=0$, $ q=\frac{1}{2}$, and eigenvalues $e_{1}(P)=0,~ e_{2}(P)=0, e_{3}(P)=-\frac{3}{2}$, $e_{4}(P)=-\frac{1}{2}$, and $e_{5}(P)=3$. The point $P$ represents a matter-dominated universe in an unstable manner.

The CP $Q$ has $w_{eff}=-1$, $ q=-1$, which means that it is a de Sitter point. At the point $Q$ the matrix of the linearized system has the following eigenvalues $e_{1}(Q)=0, e_{2}(Q)=0$, $e_{3}(Q)=-5,~e_{3}(Q)=-3$ and $e_{4}(Q)=-3$, implying that $Q$ is a non-hyperbolic point. So, the CMT is utilized to analyze its stability behavior. The details are provided in Appendix \ref{Q_stability}. It is analyzed that for $m=2$ and $n>1$, $Q$ is unstable and for $m \geq3$ and $n>1$, it has unstable nature for $h_0>0$ and stable for $h_0<0$.

The CP $R$ corresponds to a stiff-fluid decelerated universe. Its instability is evident from the eigenvalue $e_{1}(R)=0, e_{2}(R)=2$, $e_{3}(R)=3,~e_{4}(R)=6$ and $e_{5}(C)=0$.
\subsection{Analysis at infinity}
To investigate the existence of critical points at infinity, we define the
Poincare variables
\begin{equation*}
    x=\frac{X}{\rho}, ~~y=\frac{Y}{\rho},~~z=\frac{Z}{\rho}.
\end{equation*}
 where $\rho= \sqrt{1-X^2-Y^2-Z^2}$. We define the new independent variable\\ $dt=\sqrt{1-X^2-Y^2-Z^2} dT$, where $X^2 \leq 1$, $Y^2 \leq 1$, and $Z^2 \leq 1$ with constraint $1-X^2-Y^2-Z^2 \geq0$.
At infinity, the dynamical system (\ref{xeq_3})-(\ref{zetaeq_3}) becomes\footnote{The symbolic notations $E_1,~E_2,~E_3,~E_4$ and $E_5$ are used; detailed calculations are provided in Appendix \ref{conn3Infty}.}
\begin{equation}
\begin{aligned}
\frac{dX}{dT}=
E_1(X,Y,Z,\lambda,\zeta;m,n,h_0,\alpha,\beta),
\end{aligned}
\end{equation}
\begin{equation}
\begin{aligned}
\frac{dY}{dT}=
E_2(X,Y,Z,\lambda,\zeta;m,n,h_0,\alpha,\beta),
\end{aligned}
\end{equation}
\begin{equation}
\begin{aligned}
\frac{dZ}{dT}=
E_3(X,Y,Z,\lambda,\zeta;m,n,h_0,\alpha,\beta),
\end{aligned}
\end{equation}
\begin{align}
 \frac{d \lambda}{dT}=E_4(X,Y,Z,\lambda,\zeta;m,n,h_0,\alpha,\beta),
\end{align}
\begin{align}
 \frac{d \zeta}{dT}=E_5(X,Y,Z,\lambda,\zeta;m,n,h_0,\alpha,\beta).
\end{align}

The $w_{eff}$ and $q$ in terms of new variables are rewritten as
\begin{align}
    w_{eff}=-\frac{-6h_0X^2+\sqrt{6} X \zeta(Z-4\rho) +3Y\rho}{6\rho^2},
\end{align}
\begin{align}\label{Qef_conlll}
   q=\frac{6h_0X^2-\sqrt{6} X \zeta (Z-4\rho)+\rho(-3Y+2\rho)}{4\rho^2}. 
\end{align}
\begin{table}[tbp]
    \centering
    \begin{tabular}{|c|c|c|c|c|c|}
        \hline
        \textbf{Critical point}  &$(X,Y,Z,\lambda,\zeta)$ &\textbf{Existence} &\textbf{$w_{\text{eff}}$} & $q$ & \textbf{Stability}\\ \hline
       $C_{1\pm}$ & $\left( 0,\pm \sqrt{1-Z^2},Z,\lambda,\zeta \right)$ & $-1 \leq Z \leq1$ &$\pm \infty$&$\pm \infty$& unstable\\ \hline
            $C_{2\pm}$ & $\left( 0,0,\pm 1,\lambda,\zeta \right)$ & $\lambda, \zeta~$arbitrary &$0$&$\frac{1}{2}$& unstable\\ \hline
              $C_{3\pm}$ & $\left( X,0,\pm \sqrt{1-X^2},0,0 \right)$ & $-1 \leq X \leq1$ &$\infty$&$ \infty$& unstable\\ \hline
                $C_{4}$ & $\left( 1,0,0,0,0 \right)$ &Always &$ \infty$&$ \infty$& unstable\\ \hline
             \end{tabular}
     \caption{Critical points and their physical properties.}
    \label{table3_conlll}
\end{table}
\begin{figure}[tbp]
    \centering
    \begin{subfigure}{0.45\textwidth}
        \centering
        \includegraphics[width=\linewidth]{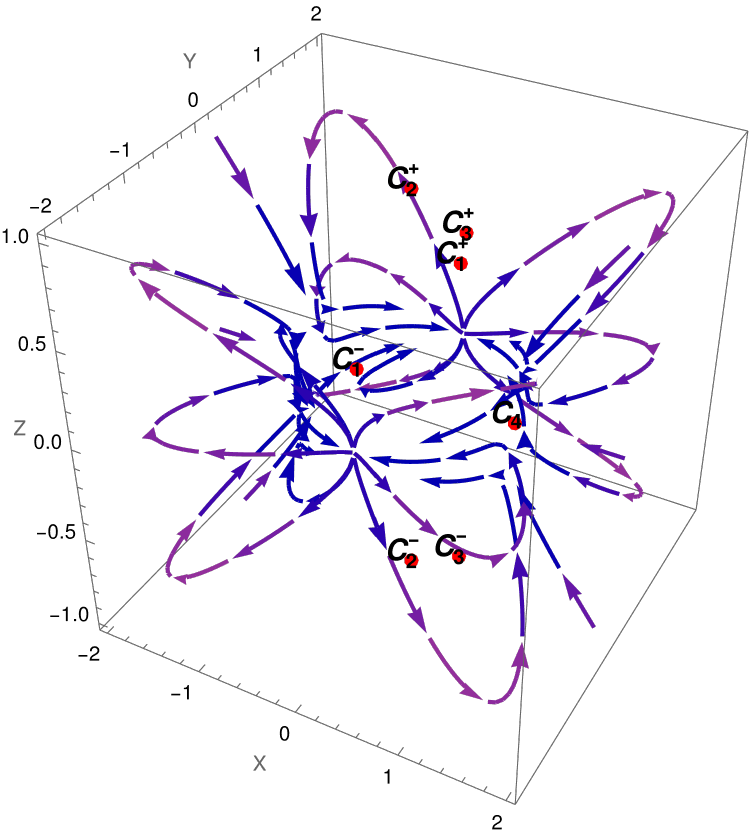}
        \caption{3D phase portrait for $h_0=0.5,~m=2,~n=2$.}
    \end{subfigure}
    \hfill
    \begin{subfigure}{0.45\textwidth}
        \centering
        \includegraphics[width=\linewidth]{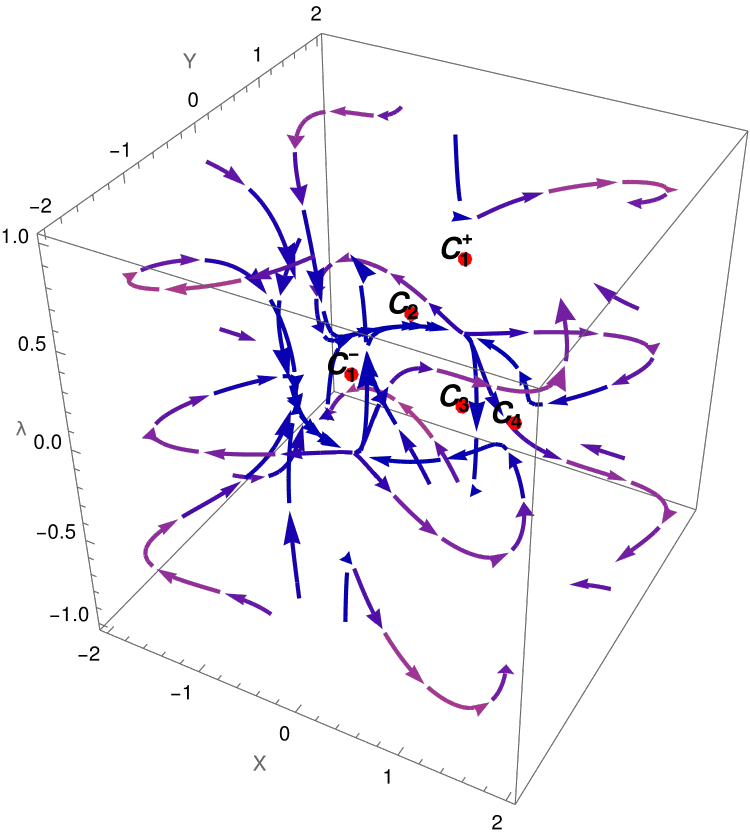}
        \caption{3D phase portrait for $h_0=0.5,~m=2,~n=2,~\alpha=1,\beta=2$.}
    \end{subfigure}

    \caption{3D phase portraits for the dynamical system given in Eqs.~(\ref{Xeq_conlll})-(\ref{Zetaeq_conlll}).}
    \label{fig1_conlll}
\end{figure}
\begin{figure}[tbp]
    \centering
    \begin{subfigure}[b]{0.45\textwidth}
        \centering
        \includegraphics[width=\textwidth]{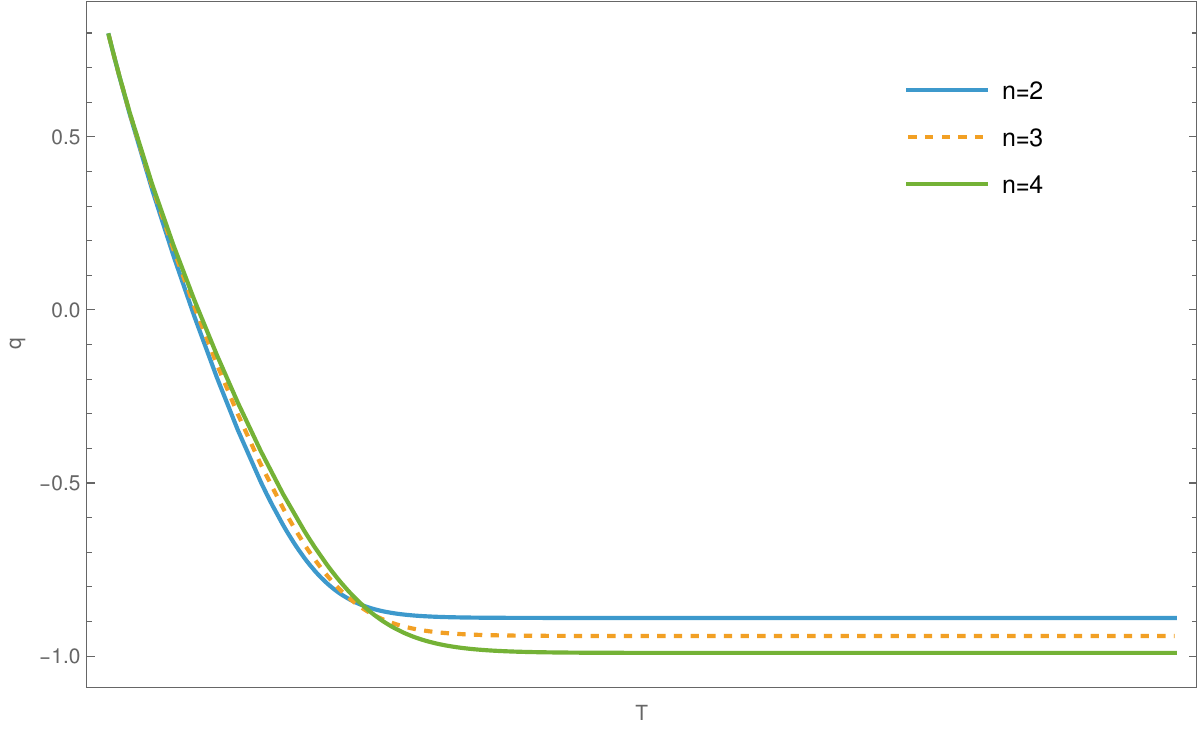}
    \end{subfigure}
    \begin{subfigure}[b]{0.45\textwidth}
        \centering
        \includegraphics[width=\textwidth]{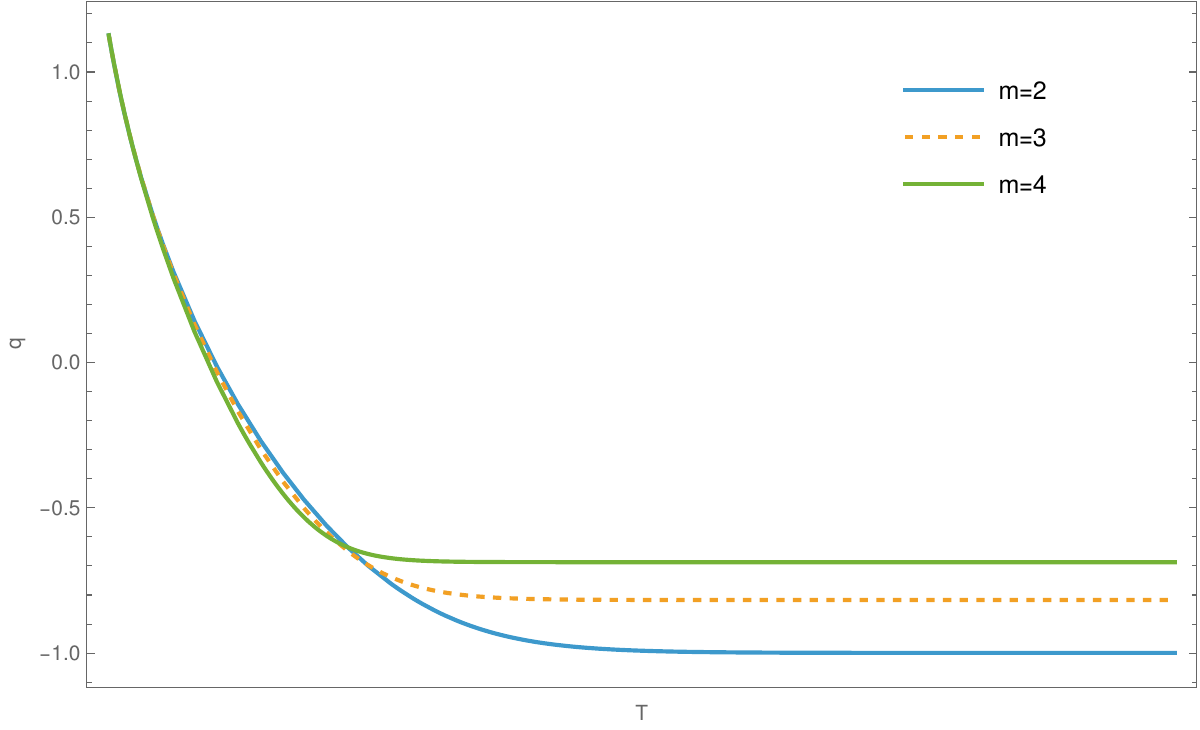}
    \end{subfigure}
    
    \caption{Qualitative evolution of the deceleration parameter of the dynamical system (\ref{Xeq_conlll})-(\ref{Zetaeq_conlll})
for different values of $n$ and $m$, with initial conditions ($X[0]=0.4,~Y[0]=0.7,~Z[0]=0.4 ,~\lambda[0]=0.4,~\zeta[0]=0.5$) with parameters values are $h_0=0.5,~\beta=2,~\alpha=1$.}
    \label{fig2_conlll}
\end{figure}
The critical points of the dynamical system analysis at infinity are mentioned in the Table \ref{table3_conlll}. The analysis describes that the points $C_{1\pm}$,  $C_{3\pm}$, and $C_{4}$ provide Big Rip or Big Crunch cosmological scenarios. But, the critical points $C_{2\pm}$ give a decelerated universe dominated by matter. In Fig \ref{fig1_conlll}, the qualitative analysis of these points is displayed in 3D phase portraits, which reveal their unstable behavior. Furthermore, the qualitative evolution of the deceleration parameter (\ref{Qef_conlll}) is also illustrated in Fig \ref{fig2_conlll}.

\section{A comparative discussion}\label{comparison}

Let us briefly compare our findings with some existing studies on non-minimally coupled scalar curvature and scalar torsion theories. In \cite{Sami2012}, a non-minimally coupled scalar curvature theory was considered, where both the coupling function and the scalar field potential were chosen in power-law forms. It was found that the system admits a de Sitter solution that acts as a dynamical attractor. Nevertheless, this solution emerges only in the regime where the effective gravitational constant becomes negative. This feature signals the onset of ghost instabilities, leading to a future evolution dominated by unphysical behavior and, consequently, rendering the model cosmologically unacceptable.

On the other hand, in \cite{Skugoreva2015}, a similar setup was investigated within the framework of non-minimally coupled scalar torsion theory. By adopting analogous choices of coupling function and potential, their dynamical analysis revealed the presence of de Sitter-type solutions; however, none of these solutions turned out to be viable candidates for a sustainable late-time de Sitter phase, as required for consistent cosmological evolution.

In contrast, our analysis shows the existence of de Sitter attractor stable late-time solutions under viable parametric conditions across all connections for steep exponential potential. This highlights that while previous works within both scalar curvature and scalar torsion frameworks encounter critical limitations, either lacking sustainable de Sitter phases or being plagued by ghost instabilities, our non-minimally coupled non-metricity scalar tensor theory is not only able to achieve stable late-time acceleration under physically consistent conditions and also gives a viable cosmological history from early to late time.

In \cite{murtazascalar}, the authors investigated non-metricity scalar-tensor theory for the second connection branch by adopting a quadratic power-law coupling function together with a simple exponential potential. Their dynamical analysis revealed the presence of a stiff-fluid solution, a matter-dominated solution, and, importantly, a stable late-time de Sitter attractor. These results are broadly consistent with our analysis, with one key distinction. In our case, for the same quadratic power-law coupling function but with a steeper potential, the de Sitter solution does not appear as a stable attractor. Instead, we find that stability of the de Sitter solution is restored when the coupling function is taken to be of higher-than-quadratic power. 

Similarly, in \cite{De2025}, the third connection branch of no-metricity scalar–tensor theory was explored within a comparable setup. Their analysis reported the existence of a stable late-time de Sitter solution. In contrast, our results show that, under the same quadratic power-law coupling, the de Sitter solution does not emerge as a stable attractor. However, we find that stability can be achieved when the coupling function is extended beyond the quadratic form. In particular, the de Sitter solution becomes stable for $h_0<0$ and unstable for $h_0>0$. This distinction underlines the strong sensitivity of late-time cosmological dynamics to both the power of the coupling function and the steepness of the potential. It further emphasizes that, within non-metricity scalar-tensor frameworks, the realization of a physically viable de Sitter attractor typically requires a coupling function of higher than quadratic order, together with a steep potential.

\section{Conclusion}\label{sec7}
In this work, we have investigated the cosmological dynamics of a steep potential within the framework of non-minimally coupled scalar-tensor non-metricity theory of gravity, considering a power-law form of the coupling function across all three independent connections in a spatially flat FLRW background. A comprehensive dynamical systems analysis has been carried out, both in the finite regime and at infinity, for each connection. A detailed critical point analysis has been performed, and the Center Manifold Theory has been employed to examine the stability of non-hyperbolic critical points. We began by formulating the field equations and introducing suitable dimensionless variables within the context of H-normalization, which allowed us to construct the corresponding autonomous dynamical systems for all three connections. For each critical point, we discussed the existence conditions, carried out a full stability analysis, and evaluated the effective equation of state parameter $w_{eff}$ and the deceleration parameter $q$.

For Connection I, the dynamical system analysis in the finite regime reveals three critical points. The point $A$ corresponds to an unstable matter-dominated universe. The CP $B$ is a potential-dominated solution, representing a de Sitter universe; however, its stability cannot be assessed via linear analysis due to the presence of zero eigenvalues. Applying center manifold theory, we find that for $m=2$, and $n>1$ the point is unstable, whereas for $m \geq 3$ and $n>1$ it acts as a stable de Sitter attractor when $h_0>0$ and exhibits instability for $h_0<0$. The point $C$ is kinetic-dominated, describing a stiff-fluid universe, and is inherently unstable. At the infinite regime, the critical points are  $A_{1_\pm}$ and $A_{2}$ and correspond to cosmological singularities, such as a Big Crunch or Big Rip.

For Connection II, the finite regime analysis yields three critical points. The point $P_1$ gives an unstable matter-dominated solution. The critical point $P_2$ is a potential-dominated point giving a de Sitter solution, but it is non-hyperbolic, so applying center manifold theory, we obtain that for $m=2$, and $n>1$ the point is unstable, whereas for $m \geq 3$ and $n>1$ it acts as a stable de Sitter attractor. The CP $P_3$ represents a stiff-fluid universe. The infinite regime analysis, reveals that the points $B_{1\pm}$,  $B_{3}$, and $B_{4\pm}$ correspond to extreme singularities. But, the points $B_{2\pm}$ give a matter-dominated universe.

For Connection III, the finite regime analysis also gives three critical points. The point $P$ represents a matter-dominated solution. The critical point $Q$ is a potential-dominated point giving a de Sitter solution, but it is also non-hyperbolic, so by using the center manifold theory, we analyze that for $m=2$, and $n>1$ the point is unstable, whereas for $m \geq 3$ and $n>1$ it acts as a stable de Sitter attractor when $h_0<0$ and unstable for $h_0>0$. The critical point $R$ represents a stiff-fluid universe. The infinite regime analysis describes that the critical points $C_{1\pm}$, $C_{3\pm}$, and $C_{4}$ give a Big Rip or Big Crunch. However, the points $C_{2\pm}$ exhibit a decelerated universe dominated by matter. 

In conclusion, our analysis of steep potential within the framework of scalar-tensor non-metricity gravity across all connection branches demonstrates that such potential can, under specific parametric conditions, can explain the early and late-time expansions of the universe. However, a definitive assessment requires further investigations, particularly those incorporating observational constraints, to determine whether steep potential can viably account for cosmic acceleration or if it encounters significant theoretical or phenomenological limitations.

Finally, it is important to note that the present analysis focuses exclusively on background dynamical stability (i.e., the phase-space stability of critical points). While a full viability assessment requires an investigation of perturbative stability, recent advancements by Ganesh et al. \cite{ganesh} have already established the scalar cosmological perturbation framework within the scalar–tensor extension of non-metricity gravity. Their derivation of the perturbed field equations and subsequent analysis of density contrast evolution under the quasi-static approximation provide a robust foundation for future research into the perturbative sector of this framework.
\appendix\label{apendix}

\section{Stability Analysis using Center Manifold Theory}\label{app1}

Following the approaches of Jack \cite{Carr} and Perko \cite{Perko2013}, we outline the mathematical foundation of the center manifold theory. When the Jacobian matrix of an autonomous system, evaluated at a critical point, possesses one or more zero eigenvalues, the linear stability analysis becomes inconclusive. In such situations, the center manifold theory provides a powerful tool, as it allows us to reduce the dimensionality of the system in the vicinity of the critical point. Specifically, there always exists a local invariant center manifold $W^c$ passing through the critical point, on which the dynamics of the system can be projected. By analyzing the reduced system on $W^c$ one can determine the local stability properties of the original system at the critical point. 

\textbf{Theorem}: Let us consider a nonlinear autonomous system in the neighborhood of a fixed point
\begin{align}\label{xeq}
\frac{dx}{dN} &= A x + f(x,y), 
\end{align}
\begin{align}\label{yeq}
\frac{dy}{dN} &= B y + g(x,y).
\end{align}
where $(x,y) \in \mathbb{R}^c \times \mathbb{R}^s$, with $c$ the dimension of $\mathbb{E}^c$ and $s$ the dimension of $\mathbb{E}^s$, and the two functions $f$ and $g$ satisfy
\begin{align}
f(0,0) &= 0,~g(0,0) = 0,\\
Df(0,0) &= 0,~Dg(0,0) = 0.
\end{align}

In the system (\ref{xeq}-\ref{yeq}), $A$ is a matrix of $c \times c$ order and represents eigenvalues having zero real parts, while $B$ is a matrix of $s \times s$ order and denotes the eigenvalues having negative real parts.

\medskip

\textbf{Center Manifold} : 
A geometrical space is a center manifold for (\ref{xeq}-\ref{yeq}), if it can be locally represented as
\begin{equation}
W^c(0) = \{ (x,y) \in \mathbb{R}^c \times \mathbb{R}^s : y = h(x), \, |x|<\delta , \, h(0)=0, \, Dh(0)=0 \},
\end{equation}
for $\delta$ sufficiently small and $h(x)$ a (sufficiently regular) function on $\mathbb{R}^s$.

\medskip

The conditions $h(0) = 0$ and $Dh(0) = 0$ from the definition imply that the space $W^c(0)$ is
tangent to the eigenspace $E^c$ at the critical point $(x,y) = (0,0)$.

\textbf{Existence}
There exists a center manifold associated with system (\ref{xeq}-\ref{yeq}). The dynamics of (\ref{xeq}-\ref{yeq}), when restricted to this center manifold, is described by
\begin{equation}\label{ueq}
\frac{du}{dN} = A u + f(u,h(u)), \quad u \in \mathbb{R}^c \ \text{ sufficiently small}.
\end{equation}

\textbf{Theorem (Stability)}:
If the zero solution of (\ref{ueq}) is stable (whether stable, asymptotically stable, or unstable), then the zero solution of (\ref{xeq}-\ref{yeq}) exhibits the same type of stability. In addition, whenever $(x(N),y(N))$ is a solution of (\ref{xeq}-\ref{yeq}) with $(x(0),y(0))$ sufficiently small, there exists a corresponding solution $u(N)$ of (\ref{ueq}) such that
\begin{align}
x(N) &= u(N) + \mathcal{O}(e^{-\eta N}), \\
y(N) &= h(u(N)) + \mathcal{O}(e^{-\eta N}),
\end{align}
as $N \to \infty$, where $\eta > 0$ is a constant.

where we also used the fact that $y = h(x)$,  must satisfy the quasilinear partial differential equation
\begin{equation}\label{quasi}
\mathcal{N}(h(x)) := Dh(x)[A x + f(x,h(x))] - B h(x) - g(x,h(x)) = 0,
\end{equation}
for it to be the center manifold.  

\subsection{Stability analysis of critical point \texorpdfstring{$B$}{B}}\label{B_stability}
 We apply the center manifold theory to study the dynamics of the system (\ref{xeq_conl})-(\ref{zetaeq_conl}) near a point $B(0,2,0,0)$. Firstly,we translate the point $B$ to origin via a transformation $x \to x,~y \to y-2,~\lambda \to \lambda, \zeta \to \zeta$ under which the system (\ref{xeq_conl})-(\ref{zetaeq_conl})  becomes
\begin{equation}
\overline{x}=\frac{\Big(-\left((\sqrt{6}\lambda+3h_0x)(y+2)\right)+2(-1+h_0x^2)(3h_0x+\sqrt{6}\zeta)\Big)}{4h_0},
\end{equation}
\begin{equation}
\overline{y}=-\frac{(y+2)\Big(-6h_0x^2+3y-2\sqrt{6}x(\lambda+\zeta)\Big)}{2},
\end{equation}
\begin{align}
 \overline{\lambda}=\lambda\left(\sqrt{6}x(n-1)(n\alpha)^{\frac{2-m}{2n-2+m}}\beta^{\frac{n}{2n-2+m}}\lambda^{\frac{m-2}{2n-2+m}}+\frac{\sqrt{3}x \zeta}{\sqrt{2}}\right),
\end{align}
\begin{align}
 \overline{\zeta}=\frac{\sqrt{3}(m-2)x\zeta^2}{m\sqrt{2}}.
\end{align}
which can be written as 
\begin{align}
\begin{pmatrix}
\dfrac{dx}{dN} \\
\dfrac{dy}{dN} \\
\dfrac{d\lambda}{dN}\\
\dfrac{d\zeta}{dN}
\end{pmatrix}
=
\begin{pmatrix}
-3 & 0 & -\frac{\sqrt{\frac{3}{2}}}{h_0} & -\frac{\sqrt{\frac{3}{2}}}{h_0}\\
0 & -3 & 0&0 \\
0 & 0 & 0&0\\
0 & 0 & 0&0
\end{pmatrix}
\begin{pmatrix}
x \\ y \\ \lambda \\ \zeta
\end{pmatrix}
+
\begin{pmatrix}
g_1 \\ g_2 \\ f_1 \\ f_2
\end{pmatrix}
\end{align}
which is not in standard form as given in (\ref{xeq})-(\ref{yeq}). Now, to bring the system into standard form, we initially find the eigenvectors corresponding to the Jacobian matrix and construct a matrix, say $S$ which is
 a matrix of the eigenvectors. The corresponding eigenvectors are given as $[1,0,0,0]^T,~[0,1,0,0]^T,~[-\frac{1}{\sqrt{6}h_0},0,1,0]^T, ~[-\frac{1}{\sqrt{6}h_0},0,0,1]^T$. The matrix $S$ and its inverse $S^{-1}$ are given below

\begin{align}
S=
\begin{pmatrix}
1 & 0 & -\frac{1}{\sqrt{6}h_0} & -\frac{1}{\sqrt{6}h_0} \\
0 & 1 & 0&0 \\
0 & 0 & 1&0\\
0 & 0 & 0&1
\end{pmatrix},~~~~~
S^{-1}=
\begin{pmatrix}
1 & 0 & \frac{1}{\sqrt{6}h_0} & \frac{1}{\sqrt{6}h_0} \\
0 & 1 & 0&0 \\
0 & 0 & 1&0\\
0 & 0 & 0&1
\end{pmatrix}
\end{align}
So, by applying the following transformation
\begin{align}
\begin{pmatrix}
\dfrac{dx}{dN} \\
\dfrac{dy}{dN} \\
\dfrac{d\lambda}{dN}\\
\dfrac{d\zeta}{dN}
\end{pmatrix}
=
S^{-1}
\begin{pmatrix}
\dfrac{dx}{dN} \\
\dfrac{dy}{dN} \\
\dfrac{d\lambda}{dN}\\
\dfrac{d\zeta}{dN}
\end{pmatrix}
\end{align}
we get the following standard form
\begin{align}
\begin{pmatrix}
\dfrac{dx}{dN} \\
\dfrac{dy}{dN} \\
\dfrac{d\lambda}{dN}\\
\dfrac{d\zeta}{dN}
\end{pmatrix}
=
\begin{pmatrix}
-3 & 0 & 0 & 0\\
0 & -3 & 0&0 \\
0 & 0 & 0&0\\
0 & 0 & 0&0
\end{pmatrix}
\begin{pmatrix}
x \\ y \\ \lambda \\ \zeta
\end{pmatrix}
+
\begin{pmatrix}
g_1 \\ g_2 \\ f_1 \\ f_2
\end{pmatrix}
\end{align}

We introduce the center manifold coordinates in the following form
\begin{equation}
\begin{aligned}
    h_{1}(\lambda,\zeta) &= a\lambda^2 + b\lambda\zeta + c\zeta^2+\mathcal{O}(\lambda,\zeta), \\
    h_{2}(\lambda,\zeta) &= d\lambda^2 + e\lambda\zeta + j\zeta^2+\mathcal{O}(\lambda,\zeta)
\end{aligned}
\end{equation}
Here to compute the coefficients, we have choose the following values:\\
case I: $m=2$ and $n>1$ treated to be generic.\\ 
case II: $m\geq 3$ and $n>1$ treated to be generic.\\ 
$$\textbf{case I: $m=2$ and $n>1$ treated to be generic}$$
By solving Eq. (\ref{quasi}) and on comparing like powers of $\lambda^2$, $\lambda \zeta$ and $\zeta^2$, we obtain the following coefficients values,
$a=\frac{\sqrt{\beta}(1-n)}{3\sqrt{6}h_0^2}, ~b=\frac{\sqrt{\beta}(1-n)}{3\sqrt{6}h_0^2}, c=0, d=-\frac{1}{3h_0},~e=-\frac{2}{3h_0}, ~ j=-\frac{1}{3h_0}$.\\
Under these, we can rewrite the center manifold coordinates in the form
\begin{equation}
\begin{aligned}
    h_{1}(\lambda,\zeta) &= \frac{\sqrt{\beta}(1-n)}{3\sqrt{6}h_0^2}\lambda^2 + \frac{\sqrt{\beta}(1-n)}{3\sqrt{6}h_0^2}\lambda\zeta+\mathcal{O}(\lambda,\zeta), \\
    h_{2}(\lambda,\zeta) &= -\frac{1}{3h_0}\lambda^2  -\frac{2}{3h_0}\lambda\zeta -\frac{1}{3h_0}\zeta^2+\mathcal{O}(\lambda,\zeta).
\end{aligned}
\end{equation}
 Finally, the dynamics of the local center manifold are determined by the Eq. (\ref{ueq}), which is simply in our case is given as
 \begin{align}
     \frac{d \lambda}{dN}&=\frac{\sqrt{\beta}(1-n)}{h_0} \lambda^2+ higher~ order~ terms\\
     \frac{d \zeta}{dN}&=0.
 \end{align}
Although the center $\zeta$ direction is neutral, however the center direction of $\lambda$ is an even parity order term, implying the unstable nature of the point $B$.\\
$$\textbf{case II: $m \geq 3$ and $n>1$ treated to be generic}$$
By solving Eq. (\ref{quasi}) and on comparing like powers of $\lambda^2$, $\lambda \zeta$ and $\zeta^2$, we obtain the following coefficients values,
$a=0, ~b=0, c=0, d=-\frac{1}{3h_0},~e=-\frac{2}{3h_0}, ~ j=-\frac{1}{3h_0}$.\\
Finally, the dynamics of the local center manifold are determined by the Eq. (\ref{ueq}), which is simply in our case is given as
 \begin{align}
     \frac{d \lambda}{dN}&=\frac{(1-n)(n \alpha)^{-\frac{1}{1+2n}}\beta^{\frac{n}{1+2n}}\lambda^{3}}{h_0}+ higher~ order~ terms\\
     \frac{d \zeta}{dN}&=-\frac{1}{6h_0} \zeta^3+ higher~ order~ terms
 \end{align}
since $n>1$, so ultimately stability depends on $h_0$, if $h_0>0$, point $B$ is stable, and for $h_0<0$, it has unstable behavior.

\subsection{Stability analysis of critical point \texorpdfstring{$P_2$}{P2}}\label{P2_stability}
 We apply the center manifold theory to study the dynamics of the system (\ref{x_conll})-(\ref{zeta_conll}) near a point $P_2( 0,2,z_c,\frac{1}{2}(\sqrt{6}z_c-2\zeta_c),\zeta_c)$. Shifting the point $P_2$ to origin via a transformation $x \to x,~y \to y-2,~z \to z-z_c,~\lambda \to \lambda-(\frac{\sqrt{3}}{\sqrt{2}}z_c-\zeta_c), \zeta \to \zeta-\zeta_c$ under which the system (\ref{x_conll})-(\ref{zeta_conll})  becomes

\begin{align}
\overline{x}=\frac{1}{4}x \left( -6 + 6 h_0 x^{2} - 3(2+y) 
- 6x\left(z+z_{c}\right) 
- 2\sqrt{6}\,x\left(\zeta+\zeta_{c}\right) 
+ \frac{4\sqrt{6}\,x\left(\zeta+\zeta_{c}\right)}{m} \right),
\end{align}
\begin{align}
\overline{y}=\frac{1}{2}(2+y)\left(6 h_0 x^{2} - 3y - 6xz + 2\sqrt{6}\,x(\lambda + \zeta)\right),
\end{align}
\begin{align}
\begin{split}
\overline{z}=&\frac{1}{4}\Bigg(
-6\,(z+z_{c}) 
+ 6h x^{2}(z+z_{c}) 
- 3(2+y)(z+z_{c}) 
- 6x\,(z+z_{c})^{2} \\
&\qquad+ \sqrt{6}(2+y)\big(2\lambda + \sqrt{6}z_{c} - 2\zeta_{c}\big)
+ 4\sqrt{6}\,(\zeta+\zeta_{c}) - \frac{8\sqrt{6}\,h(m-1)x^{2}(\zeta+\zeta_{c})}{m} 
\\&\qquad+ 2\sqrt{6}\,x(z+z_{c})(\zeta+\zeta_{c})
+\frac{4\sqrt{6}(m-1)x(z+z_{c})(\zeta+\zeta_{c})}{m}
\Bigg),
\end{split}
\end{align}
{\small\begin{align}
 \overline{\lambda}=\big(\lambda + \sqrt{\tfrac{3}{2}}\,z_{c} - \zeta_{c}\big)
\left(
\sqrt{6}(n-1)x\,(n\alpha)^{\tfrac{2-m}{-2+m+2n}}
\,\beta^{\tfrac{n}{-2+m+2n}}
\big(\lambda + \sqrt{\tfrac{3}{2}}\,z_{c} - \zeta_{c}\big)^{\tfrac{-2+m}{-2+m+2n}}
+ \sqrt{\tfrac{3}{2}}\,x(\zeta+\zeta_{c})
\right),
\end{align}}
\begin{align}
 \overline{\zeta}=\frac{\sqrt{\tfrac{3}{2}}\,(m-2)\,x\,(\zeta+\zeta_{c})^{2}}{m}.
\end{align}
The above system can be written as
\begin{align}
\begin{pmatrix}
\dfrac{dx}{dN} \\
\dfrac{dy}{dN} \\
\dfrac{dz}{dN} \\
\dfrac{d\lambda}{dN}\\
\dfrac{d\zeta}{dN}
\end{pmatrix}
=
\begin{pmatrix}
-3 & 0 & 0 & 0& 0\\
0 & -3 & 0&0& 0 \\
0 & 0 & -3&0& 0\\
0 & 0 & 0&0& 0\\
0 & 0 & 0&0& 0
\end{pmatrix}
\begin{pmatrix}
x \\ y \\ z \\ \lambda \\ \zeta
\end{pmatrix}
+
\begin{pmatrix}
g_1 \\ g_2  \\ g_3 \\ f_4 \\ f_5
\end{pmatrix}
\end{align}

We define the center manifold coordinates as
\begin{equation}\label{cmt2}
\begin{aligned}
    h_{1}(\lambda,\zeta) &= a_1\lambda^2 + a_2\lambda\zeta + a_3\zeta^2+a_4\lambda^3+a_5\lambda^2 \zeta+a_6\lambda \zeta^2+a_7\zeta^3+\mathcal{O}(\lambda,\zeta), \\
    h_{2}(\lambda,\zeta) &= b_1\lambda^2 + b_2\lambda\zeta + b_3\zeta^2+b_4\lambda^3+b_5\lambda^2 \zeta+b_6\lambda \zeta^2+b_7\zeta^3+\mathcal{O}(\lambda,\zeta), \\
    h_{3}(\lambda,\zeta) &= c_1\lambda^2 + c_2\lambda\zeta + c_3\zeta^2+c_4\lambda^3+c_5\lambda^2 \zeta+c_6\lambda \zeta^2+c_7\zeta^3+\mathcal{O}(\lambda,\zeta).
\end{aligned}
\end{equation}
Here to compute the coefficients, we have choose the following values:\\
case I: $m=2$ and $n>1$ treated to be generic.\\ 
case II: $m\geq 3$ and $n>1$ treated to be generic.\\ 
$$\textbf{case I: $m=2$ and $n>1$ treated to be generic}$$
By solving Eq. (\ref{quasi}) and on comparing like powers of $\lambda^2$, $\lambda \zeta,~\zeta^2,~\lambda^3,~\lambda^2 \zeta,~\lambda \zeta^2$ and $\zeta^3$, we obtain the following coefficients values,
$a_1=0,~a_2=0,~a_3=0~,a_4=\frac{-2\sqrt{2\beta}(1-n)}{\sqrt{3}},~a_5=0,~a_6=0,~a_7=\frac{-\sqrt{2\beta}(1-n)}{\sqrt{3}},~b_1=0,~b_2=0,~b_3=0,~b_4=0,~b_5=0,~b_6=-\frac{(1-n)}{3},~b_7=0,~ c_1=0,~ c_2=0,~ c_3=0,~ c_4=0,~ c_5=0,~ c_6=0,~ c_7=0$.\\
Under these, we can rewrite the center manifold coordinates in the form
\begin{equation}
\begin{aligned}
    h_{1}(\lambda,\zeta) &= -\frac{2\sqrt{2\beta}(1-n)}{\sqrt{3}}\lambda^3 -\frac{\sqrt{2\beta}(1-n)}{\sqrt{3}}\zeta^3+\mathcal{O}(\lambda,\zeta), \\
    h_{2}(\lambda,\zeta) &= \frac{(n-1)}{3}\lambda \zeta^2 +\mathcal{O}(\lambda,\zeta).\\
    h_{3}(\lambda,\zeta) &=0+\mathcal{O}(\lambda,\zeta).
\end{aligned}
\end{equation}
 So, the dynamics of the local center manifold are given as
 \begin{align}
     \frac{d \lambda}{dN}&=(4\beta-8n\beta+4n^2 \beta) \lambda^4+ higher~ order~ terms\\
     \frac{d \zeta}{dN}&=0.
 \end{align}
Although the center $\zeta$ direction is neutral, however the center direction of $\lambda$ is an even parity order term, implying the unstable nature of the point $P_2$. Also, it is important to note that if $n=1$, then $\frac{d \lambda}{dN}$ also reduces to zero, and we cannot determine stability; instead, we must consider higher-order terms.\\
$$\textbf{case II: $m \geq 3$ and $n>1$ treated to be generic}$$
For this case, we  get the following values of the coefficients,
$a_1=0,~a_2=0,~a_3=0~,a_4=\frac{1-n}{\sqrt{3}},~a_5=0,~a_6=0,~a_7=0,~b_1=0,~b_2=0,~b_3=0,~b_4=0,~b_5=0,~b_6=0,~b_7=0,~ c_1=0,~ c_2=0,~ c_3=0,~ c_4=0,~ c_5=0,~ c_6=0,~ c_7=0$.\\\\
Under these we can rewrite the center manifold coordinates in the form
\begin{equation}
\begin{aligned}
    h_{1}(\lambda,\zeta) &= \frac{1-n}{\sqrt{3}}\lambda^3 +\mathcal{O}(\lambda,\zeta), \\
    h_{2}(\lambda,\zeta) &= 0 +\mathcal{O}(\lambda,\zeta).\\
    h_{3}(\lambda,\zeta) &=0+\mathcal{O}(\lambda,\zeta).
\end{aligned}
\end{equation}
 So, the dynamics of the local center manifold are given as
 \begin{align}
     \frac{d \lambda}{dN}&=(\frac{1-n}{\sqrt{2}}) \lambda^4 \zeta+ higher~ order~ terms\\
     \frac{d \zeta}{dN}&=(\frac{1-n}{3\sqrt{2}}) \lambda^3 \zeta^2+ higher~ order~ terms.
 \end{align}
Here the lowest power term of the expression of the center manifold depends
only on $\zeta$,  the nature of the vector field implicitly depends on $\lambda$, not explicitly. Now we try to write the flow in terms of $\zeta$ only.
\begin{align*}
    \frac{d \lambda}{d \zeta}=\frac{3\lambda}{\zeta}
\end{align*}
implies that,
\begin{align*}
 \lambda=C \zeta^3,~ where~ C ~is~positive~arbitrary~constant. 
\end{align*}
this substitution gives,
 \begin{align}
     \frac{d \zeta}{dN}&=(\frac{1-n}{3\sqrt{2}})C^3 \zeta^{11}.
 \end{align}
Since $n>1$, implies that $P_2$ is stable.\\

\subsection{Stability analysis of critical point \texorpdfstring{$Q$}{Q}}\label{Q_stability}
 We apply the center manifold theory to study the dynamics of the system (\ref{xeq_3})-(\ref{zetaeq_3}) near a point $Q( 0,2,z_c,\frac{1}{2}(3z_c-2)\zeta_c,\zeta_c)$. Shifting the point $Q$ to origin via a transformation $x \to x,~y \to y-2,~z \to z-z_c,~\lambda \to \lambda-\frac{1}{2}(3z_c-2)\zeta_c, \zeta \to \zeta-\zeta_c$ under which the system (\ref{xeq_3})-(\ref{zetaeq_3})  becomes
\begin{align}
\overline{x}=&\frac{1}{4 \left( 4 \sqrt{6}\, h x + 3 (z + z_{c})(\zeta + \zeta_{c}) \right)} \; 3x \Bigg( \nonumber 8 \sqrt{6}\, h^{2} x^{3}  - 2 h x^{2} (-8 + z + z_{c})(\zeta + \zeta_{c}) \nonumber \\
&\quad + \big( -16 - (-4 + 3y)(z + z_{c}) \big)(\zeta + \zeta_{c})  - 4(2 + y)\big( 2 \lambda + (-2 + 3 z_{c})\zeta_{c} \big) \nonumber \\
&\quad - \sqrt{6} x \Big( 4h(4 + y) 
      + (z + z_{c})(2 - \tfrac{4}{m} + z + z_{c})(\zeta + \zeta_{c})^{2} \Big)
\Bigg),
\end{align}
\begin{align}
\overline{y}=\frac{1}{2}(2 + y)\Big(& 
    6 + 6 h x^{2} - 3(2 + y) - \sqrt{6}\, x \Big( -2 \lambda - 2 \zeta 
        + z \zeta 
        + z_{c}(\zeta - 2 \zeta_{c}) 
        + z \zeta_{c} \Big)
\Big),
\end{align}
\begin{align}
\overline{z}=&\frac{1}{4 \left( 4 \sqrt{6}\, h x + 3 (z + z_{c})(\zeta + \zeta_{c}) \right)}
(z + z_{c}) \Big(
24 \sqrt{6}\, h^{2} x^{3}- 6 h x^{2} \big(-8 - \tfrac{8}{m} + z + z_{c}\big)(\zeta + \zeta_{c}) \nonumber \\
    &- 3 \big(-8 + 3 (4 + y)(z + z_{c})\big)(\zeta + \zeta_{c})+ 6 (2 + y)\big( 2 \lambda + (-2 + 3 z_{c}) \zeta_{c} \big) \nonumber \\
    &+ \sqrt{6}\, x \Big( 8h - 12h(2 + y) 
      - 3(-4 + z + z_{c})(z + z_{c})(\zeta + \zeta_{c})^{2} \Big)
\Big),
\end{align}

\begin{align}
\resizebox{\textwidth}{!}{$
\overline{\lambda} =
\Big(\lambda + \Big(-1 + \tfrac{3 z_{c}}{2}\Big)\zeta_{c}\Big)
\Big(
    \sqrt{\tfrac{3}{2}}\, x (\zeta + \zeta_{c})
    + \sqrt{6}\, (-1 + n)\, x\,
       (n \alpha)^{\tfrac{2 - m}{-2 + m + 2n}}\,
       \beta^{\tfrac{n}{-2 + m + 2n}}\,
       \Big(\lambda + \Big(-1 + \tfrac{3 z_{c}}{2}\Big)\zeta_{c}\Big)^{\tfrac{-2 + m}{-2 + m + 2n}}
\Big)
$}
\end{align}
\begin{align}
 \overline{\zeta}=\frac{\sqrt{\tfrac{3}{2}}\,(m-2)\,x\,(\zeta+\zeta_{c})^{2}}{m}.
\end{align}

The above system can be written as
 \begin{align}
\begin{pmatrix}
\dfrac{dx}{dN} \\[6pt]
\dfrac{dy}{dN} \\[6pt]
\dfrac{dz}{dN} \\[6pt]
\dfrac{d\lambda}{dN} \\[6pt]
\dfrac{d\zeta}{dN}
\end{pmatrix}
=
\begin{pmatrix}
-5 & 0 & 0 & 0 & 0\\
0 & -3 & 0 & 0 & 0\\
0 & 0 & -3 & 0 & 0\\
0 & 0 & 0 & 0 & 0\\
0 & 0 & 0 & 0 & 0
\end{pmatrix}
\begin{pmatrix}
x \\[4pt]
y \\[4pt]
z \\[4pt]
\lambda \\[4pt]
\zeta
\end{pmatrix}
+
\begin{pmatrix}
g_1 \\[4pt]
g_2 \\[4pt]
g_3 \\[4pt]
f_4 \\[4pt]
f_5
\end{pmatrix}
\end{align}

We propose the center manifold coordinates as given in (\ref{cmt2}).\\
Here to compute the coefficients, we have choose the following values:\\
case I: $m=2$ and $n>1$ treated to be generic.\\ 
case II: $m\geq 3$ and $n>1$ treated to be generic.\\ 
$$\textbf{case I: $m=2$ and $n>1$ treated to be generic}$$ 
Comparing like powers, we get the following coefficients values,
$a_1=0,~a_2=0,~a_3=0~,a_4=\frac{1}{4 \sqrt{6}h_0},~a_5=0,~a_6=0,~a_7=0,~b_1=0,~b_2=0,~b_3=0,~b_4=0,~b_5=0,~b_6=0,~b_7=0,~ c_1=0,~ c_2=0,~ c_3=0,~ c_4=0,~ c_5=0,~ c_6=0,~ c_7=0$.\\
So, the dynamics of the local center manifold are given as
 \begin{align}
     \frac{d \lambda}{dN}&=(-\frac{\sqrt{\beta}}{2\sqrt{2}h_0}+\frac{n\sqrt{\beta}}{2\sqrt{2}h_0}) \lambda^4+ higher~ order~ terms\\
     \frac{d \zeta}{dN}&=0.
 \end{align}
Although the center $\zeta$ direction is neutral, the center direction of $\lambda$ is an even parity order term, implying the unstable nature of the point $Q$.\\
$$\textbf{case II: $m \geq 3$ and $n>1$ treated to be generic}$$
For this case, we  get the following values of the coefficients,
$a_1=0,~a_2=0,~a_3=0~,a_4=\frac{1}{4\sqrt{6}h_0},~a_5=0,~a_6=0,~a_7=0,~b_1=0,~b_2=0,~b_3=0,~b_4=0,~b_5=0,~b_6=0,~b_7=-\frac{2}{9},~ c_1=0,~ c_2=0,~ c_3=0,~ c_4=0,~ c_5=0,~ c_6=0,~ c_7=0$.\\\\
So, the dynamics of the local center manifold are given as
 \begin{align}
     \frac{d \lambda}{dN}&=\frac{\lambda^4 \zeta}{8h_0} + higher~ order~ terms\\
     \frac{d \zeta}{dN}&=\frac{\lambda^3 \zeta^2}{24h_0} + higher~ order~ terms.
 \end{align}
Here the lowest power term of the expression of the center manifold depends
only on $\zeta$,  the nature of the vector field implicitly depends on $\lambda$, not explicitly. Now we try to write the flow in terms of $\zeta$ only.
\begin{align*}
    \frac{d \lambda}{d \zeta}=\frac{3\lambda}{\zeta}
\end{align*}
implies that,
\begin{align*}
 \lambda=C \zeta^3,~ where~ C ~is~positive~arbitrary~constant. 
\end{align*}
this substitution gives,
 \begin{align}
     \frac{d \zeta}{dN}&=\frac{C^3 \zeta^{11}}{24h_0}.
 \end{align}
for $h_0>0$, $Q$ is unstable, and for $h_0<0$, it is stable.

\section{Dynamical System Equations at infinity for Connection I}\label{conn1Infty}
The cosmological field equations for the Connection I, within the compactified dimensionless variables are as follows.
{\small\begin{equation}\label{Xeq_conl}
\begin{aligned}
\frac{dX}{dT} =
& -\frac{1}{4h_0 } \Bigg(
  6h(1+h_0)X^5+ 3h_0X^3 \Big(-2(2+h_0) + (-2+4h_0)Y^2 - Y\sqrt{1 - X^2 - Y^2}\Big) \\
&+ 3h_0X\Big(2 + 2Y^2 - 4Y^4 + Y\sqrt{1 - X^2 - Y^2}
   - 2Y^3 \sqrt{1 - X^2 - Y^2}\Big) \\
&- \sqrt{6}(1 - Y^2)\Big( Y\lambda + 2\sqrt{1 - X^2 - Y^2}\,\zeta \Big)+ \sqrt{6}X^4\Big( Y\lambda + 2(1+h_0)\sqrt{1 - X^2 - Y^2}\,\zeta \Big) \\
&+ \sqrt{6}X^2\Big( -2Y\lambda + Y^3\lambda
   - 2(2+h_0)\sqrt{1 - X^2 - Y^2}\,\zeta + 2Y^2\sqrt{1 - X^2 - Y^2}(2h_0\lambda + \zeta + 2h_0\zeta) \Big)
\Bigg),
\end{aligned}
\end{equation}}
{\small\begin{equation}
\begin{aligned}
\frac{dY}{dT} = 
& -\frac{1}{4h_0} \, Y \Bigg(
   6h(1+h_0)X^4+ 3h_0X^2 \Big(2 - 4h_0 + (-2+4h_0)Y^2 - Y\sqrt{1 - X^2 - Y^2}\Big) \\
&- 6h_0(-1+Y^2)\Big(-2 + 2Y^2 + Y\sqrt{1 - X^2 - Y^2}\Big) + \sqrt{6}\,X^3\Big(Y\lambda + 2(1+h_0)\sqrt{1 - X^2 - Y^2}\,\zeta \Big) \\
&+ \sqrt{6}\,X(-1+Y^2)\Big(Y\lambda + 2\sqrt{1 - X^2 - Y^2}\,
   (2h_0\lambda + \zeta + 2h_0\zeta)\Big)
\Bigg),
\end{aligned}
\end{equation}}

\begin{align}
 \frac{d \lambda}{dT}=\lambda\left(\sqrt{6}X(n-1)(n\alpha)^{\frac{2-m}{2n-2+m}}\beta^{\frac{n}{2n-2+m}}\lambda^{\frac{m-2}{2n-2+m}}+\frac{\sqrt{3}X \zeta}{\sqrt{2}}\right),
\end{align}
\begin{align}\label{Zetaeq_conl}
 \frac{d \zeta}{dT}=\frac{\sqrt{3}(m-2)X\zeta^2}{m\sqrt{2}}.
\end{align}

\section{Dynamical System Equations at infinity for Connection II}\label{conn2Infty}
The cosmological field equations for the Connection II, within the compactified dimensionless variables, are as follows.
\begin{equation}\label{Xeq_conll}
\begin{aligned}
\frac{dX}{dT}=
\frac{X}{4 m }
& \Bigg(
   -6 (1+h_0)m X^4   + 2X^3 \left(3mZ + \sqrt{6}(-2+m)\sqrt{1 - X^2 - Y^2 - Z^2}\,\zeta \right) \\
&  + X^2 \Big( 6(2+h_0)m + 6(1-2h_0)mY^2 - 6(2+h_0)mZ^2 \\
&  + mY \Big( 3\sqrt{1 - X^2 - Y^2 - Z^2} + 2\sqrt{6}Z\lambda \Big) \\
&  + 4\sqrt{6}\big(2h_0(-1+m)+m\big)Z\sqrt{1 - X^2 - Y^2 - Z^2}\,\zeta \Big) \\
&  + m \Big( 12Y^4 
   + 2Y^3 \Big( 3\sqrt{1 - X^2 - Y^2 - Z^2} + \sqrt{6}Z\lambda \Big) \\
&  + Y(-1+Z^2)\Big( 3\sqrt{1 - X^2 - Y^2 - Z^2} + 2\sqrt{6}Z\lambda \Big) \\
&  - 2(-1+Z^2)\Big(-3+3Z^2 - 2\sqrt{6}Z\sqrt{1 - X^2 - Y^2 - Z^2}\,\zeta \Big) \\
&  + Y^2\Big(-6+6Z^2 + 4\sqrt{6}Z\sqrt{1 - X^2 - Y^2 - Z^2}\,\zeta \Big) \Big) \\
&  + 2X \Big( 3m(-1+2Y^2)Z + 3mZ^3 + \sqrt{6}(2-3m)Z^2\sqrt{1 - X^2 - Y^2 - Z^2}\,\zeta \\
&  - \sqrt{6}\sqrt{1 - X^2 - Y^2 - Z^2}\Big((-2+m)\zeta + 2mY^2(\lambda+\zeta)\Big) \Big)
\Bigg),
\end{aligned}
\end{equation}
{\small\begin{equation}
\begin{aligned}
\frac{dY}{dT}=\frac{Y}{4 m }
& \Bigg(
   -6 (1+h_0)m X^4  + 2X^3 \left(3mZ + \sqrt{6}(-2+m)\sqrt{1 - X^2 - Y^2 - Z^2}\,\zeta \right) \\
&  + X^2 \Big( -6m + 12hm + 6(1-2h_0)mY^2 - 6(2+h_0)mZ^2  \\
& + mY \Big( 3\sqrt{1 - X^2 - Y^2 - Z^2} + 2\sqrt{6}Z\lambda \Big) \\
&  + 4\sqrt{6}\big(2h_0(-1+m)+m\big)Z\sqrt{1 - X^2 - Y^2 - Z^2}\,\zeta \Big) + 2X \Big( 6m(-1+Y^2)Z + 3mZ^3 \\
& + \sqrt{6}(2-3m)Z^2\sqrt{1 - X^2 - Y^2 - Z^2}\,\zeta  - 2\sqrt{6}m(-1+Y^2)\sqrt{1 - X^2 - Y^2 - Z^2}(\lambda+\zeta) \Big) \\
&  + m \Big( 12Y^4 
   + 2Y^3 \Big( 3\sqrt{1 - X^2 - Y^2 - Z^2} + \sqrt{6}Z\lambda \Big) \\
& + Y \Big( -6\sqrt{1 - X^2 - Y^2 - Z^2} 
         + 3Z^2\sqrt{1 - X^2 - Y^2 - Z^2} - 2\sqrt{6}Z\lambda + 2\sqrt{6}Z^3\lambda \Big) \\
&  - 2(-1+Z^2)\Big( 6+3Z^2 - 2\sqrt{6}Z\sqrt{1 - X^2 - Y^2 - Z^2}\,\zeta \Big) \\
&  + Y^2\Big( -24+6Z^2 + 4\sqrt{6}Z\sqrt{1 - X^2 - Y^2 - Z^2}\,\zeta \Big) \Big)
\Bigg),
\end{aligned}
\end{equation}}
{\small\begin{equation}
\begin{aligned}
\frac{dZ}{dT}=-\frac{1}{4 m}
& \Bigg(
   6mZ^5  + Z^3 \Big( 6(2+h_0)mX^2 
     - 3m\big(4 + 2Y^2 + Y\sqrt{1 - X^2 - Y^2 - Z^2}\big) \\
&  + 2\sqrt{6}(-2+3m)X\sqrt{1 - X^2 - Y^2 - Z^2}\,\zeta \Big) + 2\sqrt{6}\Big( m(-1+X^2)Y\lambda + mY^3\lambda \\
&  + 2\big(-m+(2h_0(-1+m)+m)X^2\big)\sqrt{1 - X^2 - Y^2 - Z^2}\,\zeta   \\
& + 2mY^2\sqrt{1 - X^2 - Y^2 - Z^2}\,\zeta \Big) \\
&  - 2mZ^4\Big( 3X + \sqrt{6}\big(Y\lambda + 2\sqrt{1 - X^2 - Y^2 - Z^2}\,\zeta\big)\Big)- 2Z^2 \Big( 3mX^3 + 3mX(-1+2Y^2) \\
&  + \sqrt{6}m(-2+Y^2)\big(Y\lambda + 2\sqrt{1 - X^2 - Y^2 - Z^2}\,\zeta\big) \\
&  + \sqrt{6}X^2\Big(mY\lambda + 2(2h_0(-1+m)+m)\sqrt{1 - X^2 - Y^2 - Z^2}\,\zeta\Big) \Big) \\
& + Z \Big( 6(1+h_0)mX^4 + 3mX^2\Big(-2(2+h_0)+(-2+4h_0)Y^2 - Y\sqrt{1 - X^2 - Y^2 - Z^2}\Big) \\
&  + 3m\Big(2+2Y^2 -4Y^4 + Y\sqrt{1 - X^2 - Y^2 - Z^2} - 2Y^3\sqrt{1 - X^2 - Y^2 - Z^2}\Big) \\
&  - 2\sqrt{6}(-2+m)X^3\sqrt{1 - X^2 - Y^2 - Z^2}\,\zeta \\
&  + 2\sqrt{6}X\sqrt{1 - X^2 - Y^2 - Z^2}\Big((2-3m)\zeta + 2mY^2(\lambda+\zeta)\Big) \Big)
\Bigg),
\end{aligned}
\end{equation}}
\begin{align}
 \frac{d \lambda}{dT}=\lambda\left(\sqrt{6}X(n-1)(n\alpha)^{\frac{2-m}{2n-2+m}}\beta^{\frac{n}{2n-2+m}}\lambda^{\frac{m-2}{2n-2+m}}+\frac{\sqrt{3}X \zeta}{\sqrt{2}}\right),
\end{align}
\begin{align}\label{Zetaeq_conll}
 \frac{d \zeta}{dT}=\frac{\sqrt{3}(m-2)X\zeta^2}{m\sqrt{2}}.
\end{align}

\section{Dynamical System Equations at infinity for Connection III}\label{conn3Infty}
The cosmological field equations for Connection III, within the compactified dimensionless variables, are as follows.
{\small\begin{equation}\label{Xeq_conlll}
\begin{aligned}
\frac{dX}{dT}=
&-\Bigg(X \Bigg(
24 \sqrt{6}\, h_0 (1+h_0) m X^5 
+ 6 m X^4 \Big( 4 Y \lambda 
+ \big(- (5+h_0) Z + 8 (1+h_0) \sqrt{1 - X^2 - Y^2 - Z^2}\big) \zeta  \Big) \\[6pt]
&- 3 m \Bigg( 12 Y^4 Z \zeta 
+ 2 (-1 + Z^2)\Big(-5 Z - 3 Z^3 + 8 \sqrt{1 - X^2 - Y^2 - Z^2} 
+ 4 Z^2 \sqrt{1 - X^2 - Y^2 - Z^2}\Big)\zeta  \\[6pt]
&+ 2 Y^2\Big(-11 Z + 3 Z^3 + 8 \sqrt{1 - X^2 - Y^2 - Z^2} 
+ 4 Z^2 \sqrt{1 - X^2 - Y^2 - Z^2}\Big)\zeta  \\[6pt]
&+ Y(-1 + Z^2)\Big(4 (2+Z^2)\lambda 
+ 3 Z \sqrt{1 - X^2 - Y^2 - Z^2}\, \zeta \Big) \\
& + Y^3\Big(4 (2+Z^2)\lambda 
+ 6 Z \sqrt{1 - X^2 - Y^2 - Z^2}\, \zeta \Big) \Bigg) \\[6pt]
&+ 3 X^2 \Bigg( 8 m Y^3 \lambda 
+ 2 \Big((10+h_0)m Z - (2+h_0)m Z^3 - 8 (2+h_0)m \sqrt{1 - X^2 - Y^2 - Z^2} \\
& + 4 (m + 2 h_0 (1+m)) Z^2 \sqrt{1 - X^2 - Y^2 - Z^2}\Big)\zeta  \\
&  + m Y \Big(4(-4+Z^2)\lambda - 3 Z \sqrt{1 - X^2 - Y^2 - Z^2}\, \zeta \Big) \\[6pt]
&+ 2 m Y^2 \Big(16 h_0 \sqrt{1 - X^2 - Y^2 - Z^2}\, \lambda 
+ \big(-(11+2h_0)Z + 8(1+2h_0)\sqrt{1 - X^2 - Y^2 - Z^2}\big)\zeta  \Big) \Bigg) \\[6pt]
&+ \sqrt{6}\, X^3 \Big(-24 h_0 (2+h_0)m + 24 h_0 (-1+2h_0) m Y^2 
 \\
& - 12 h_0 m Y \sqrt{1 - X^2 - Y^2 - Z^2} + 6 (2+m) Z \sqrt{1 - X^2 - Y^2 - Z^2}\,\zeta ^2  \\[6pt]
&+ m Z^2 \big(8h_0(2+3h_0) - 3 \zeta ^2\big)\Big)- \sqrt{6}\, X \Big(-24 h_0 m + 48 h_0 m Y^4 
+ 24 h_0 m Y^3 \sqrt{1 - X^2 - Y^2 - Z^2} \\[6pt]
& + 12 h_0 m Y \sqrt{1 - X^2 - Y^2 - Z^2}(-1+Z^2) 
+ 6 (2+m) Z \sqrt{1 - X^2 - Y^2 - Z^2}\,\zeta ^2 \\[6pt]
& - 12 m Z^3 \sqrt{1 - X^2 - Y^2 - Z^2}\, \zeta ^2 
+ m Z^2(16 h_0 - 3 \zeta ^2) + m Z^4(8h_0+3 \zeta ^2)+ 2 m Y^2 \\[6pt]
&\times \Big(-12 h_0 - 6 Z \sqrt{1 - X^2 - Y^2 - Z^2}\, \zeta (\lambda+\zeta ) 
+ Z^2(28h_0+3\zeta ^2)\Big)\Big)\Bigg)\Bigg/ \Big(4m (4\sqrt{6}h_0X+3Z\zeta)\Big)\Bigg),
\end{aligned}
\end{equation}}
{\small\begin{equation}
\begin{aligned}
\frac{dY}{dT}=
Y& \Bigg(
-24 \sqrt{6}\, h_0 (1+h_0) m X^5 
- 6 m X^4 \Big(4 Y \lambda 
+ \big(- (5+h_0) Z + 8 (1+h_0) \sqrt{1 - X^2 - Y^2 - Z^2}\big)\zeta \Big) \\[6pt]
&+ 3 m Z \Big(12 Y^4 \zeta  
+ 2(-1+Z^2)\big(-6 - 3 Z^2 + 4 Z \sqrt{1 - X^2 - Y^2 - Z^2}\big)\zeta  \\[6pt]
& + 2 Y^2\big(-12 + 3 Z^2 + 4 Z \sqrt{1 - X^2 - Y^2 - Z^2}\big)\zeta  + Y^3 \big(4 Z \lambda + 6 \sqrt{1 - X^2 - Y^2 - Z^2}\,\zeta \big) \\[6pt]
& + Y\big(-4 Z \lambda + 4 Z^3 \lambda - 6 \sqrt{1 - X^2 - Y^2 - Z^2}\, \zeta  
+ 3 Z^2 \sqrt{1 - X^2 - Y^2 - Z^2}\, \zeta  \big)\Big) \\[6pt]
&- 3 X^2 \Big( 8 m Y^3 \lambda 
- 32 h_0 m \sqrt{1 - X^2 - Y^2 - Z^2}\, \lambda \\[6pt]
& + 2\big((11+2h_0)m Z - (2+h_0)m Z^3 - 8(1+2h_0)m \sqrt{1 - X^2 - Y^2 - Z^2} \\
& + 4(m+2h_0(1+m))Z^2 \sqrt{1 - X^2 - Y^2 - Z^2}\big)\zeta  \\
&  + m Y \big(4(-2+Z^2)\lambda - 3 Z \sqrt{1 - X^2 - Y^2 - Z^2}\,\zeta  \big) \\[6pt]
& + 2 m Y^2 \Big(16 h_0 \sqrt{1 - X^2 - Y^2 - Z^2}\, \lambda 
+ \big(-(11+2h_0)Z + 8(1+2h_0)\sqrt{1 - X^2 - Y^2 - Z^2}\big)\zeta  \Big)\Big) \\[6pt]
&- \sqrt{6}\, X^3 \Big(24 (1-2h_0) h_0 m + 24 h_0 (-1+2h_0) m Y^2 
- 12 h_0 m Y \sqrt{1 - X^2 - Y^2 - Z^2} \\[6pt]
&+ 6 (2+m) Z \sqrt{1 - X^2 - Y^2 - Z^2}\,\zeta ^2 
+ m Z^2 \big(8h_0(2+3h_0) - 3 \zeta ^2\big)\Big) \\[6pt]
&+ \sqrt{6}\, m X \Big(48 h_0 + 48 h_0 Y^4 
+ 24 h_0 Y^3 \sqrt{1 - X^2 - Y^2 - Z^2} \\[6pt]
&+ 12 h_0 Y \sqrt{1 - X^2 - Y^2 - Z^2}(-2+Z^2) 
- 12 Z^3 \sqrt{1 - X^2 - Y^2 - Z^2}\,\zeta ^2 \\[6pt]
&+ 12 Z \sqrt{1 - X^2 - Y^2 - Z^2}\,\zeta (\lambda+\zeta ) 
+ Z^4(8h_0+3\zeta ^2) - 2 Z^2(28h_0+3\zeta ^2) \\[6pt]
&+ 2 Y^2 \Big(-48 h_0 - 6 Z \sqrt{1 - X^2 - Y^2 - Z^2}\,\zeta (\lambda+\zeta ) 
+ Z^2(28h_0+3\zeta ^2)\Big)\Big)\Bigg)\Bigg/ \Big(4m(4\sqrt{6}h_0X+3Z\zeta)\Big),
\end{aligned}
\end{equation}}
{\small\begin{equation}
\begin{aligned}
\frac{dZ}{dT}=
-&\Bigg( Z \Big( 24 \sqrt{6}\, h_0 (1+h_0) m X^{5} 
+ 6 m X^{4} \big( 4 Y \lambda 
+ (-(5+h_0) Z + 8 (1+h_0) \sqrt{1 - X^{2} - Y^{2} - Z^{2}})\, \zeta \big) \\
&  - 3 m \Big( 12 Y^{4} Z \zeta  + 2 (-1+Z^{2})^{2} \big(-3Z + 4\sqrt{1 - X^{2} - Y^{2} - Z^{2}}\big)\zeta \\
&  + 2 Y^{2} (-1+Z^{2}) \big(3Z + 4\sqrt{1 - X^{2} - Y^{2} - Z^{2}}\big)\zeta \\
&  + Y (-1+Z^{2}) \big( 4(-1+Z^{2}) \lambda + 3Z\sqrt{1 - X^{2} - Y^{2} - Z^{2}}\zeta \big) \\
&  + Y^{3} \big( 4(-1+Z^{2}) \lambda + 6 Z\sqrt{1 - X^{2} - Y^{2} - Z^{2}} \zeta \big) \Big) \\
&  + 3 X^{2} \Big( 8 m Y^{3} \lambda 
+ 2 (-1+Z^{2}) \big( -(2+h_0)mZ + 4(m+2h_0(1+m))\sqrt{1 - X^{2} - Y^{2} - Z^{2}} \big)\zeta \\
& + m Y \big( 4(-1+Z^{2})\lambda - 3Z\sqrt{1 - X^{2} - Y^{2} - Z^{2}}\zeta \big) \\
&  + 2m Y^{2} \big(16h_0 \sqrt{1 - X^{2} - Y^{2} - Z^{2}} \lambda 
+ (-(11+2h_0)Z + 8(1+2h_0)\sqrt{1 - X^{2} - Y^{2} - Z^{2}})\zeta \big) \Big) \\
&  + \sqrt{6} X^{3} \Big( -8h_0(2+3h_0)m + 24h_0(-1+2h_0)mY^{2} - 12hmY\sqrt{1 - X^{2} - Y^{2} - Z^{2}} \\
&  + 6(2+m)Z\sqrt{1 - X^{2} - Y^{2} - Z^{2}}\zeta^{2} 
+ mZ^{2}\big(8h_0(2+3h_0) - 3\zeta^{2}\big) \Big) \\
&  - \sqrt{6} m X \Big( 48h_0 Y^{4} 
+ 24h_0 Y^{3}\sqrt{1 - X^{2} - Y^{2} - Z^{2}} 
+ 12h_0 Y \sqrt{1 - X^{2} - Y^{2} - Z^{2}}(-1+Z^{2}) \\
&  + (-1+Z^{2})\big( -8h_0 - 12Z\sqrt{1 - X^{2} - Y^{2} - Z^{2}} \zeta^{2} + Z^{2}(8h_0+3\zeta^{2}) \big) \\
&  + 2 Y^{2} \big( -28h_0 - 6Z\sqrt{1 - X^{2} - Y^{2} - Z^{2}} \zeta (\lambda+\zeta) 
+ Z^{2}(28h_0+3\zeta^{2}) \big) \Big)\Bigg/ \Big(4m(4\sqrt{6}h_0X+3Z\zeta)\Big)\Bigg),
\end{aligned}
\end{equation}}
\begin{align}
 \frac{d \lambda}{dT}=\lambda\left(\sqrt{6}X(n-1)(n\alpha)^{\frac{2-m}{2n-2+m}}\beta^{\frac{n}{2n-2+m}}\lambda^{\frac{m-2}{2n-2+m}}+\frac{\sqrt{3}X \zeta}{\sqrt{2}}\right),
\end{align}
\begin{align}\label{Zetaeq_conlll}
 \frac{d \zeta}{dT}=\frac{\sqrt{3}(m-2)X\zeta^2}{m\sqrt{2}}.
\end{align}

\end{document}